%% file: main_arXiv.tex
\documentclass[conference,10pt]{IEEEtran}

\usepackage{etex}

\usepackage[english]{babel}
\usepackage[T1]{fontenc}
\usepackage{cite,url,color} 
\usepackage{graphics,amsfonts}
\usepackage{epstopdf}
\usepackage[pdftex]{graphicx}
\usepackage[cmex10]{amsmath}
\usepackage[utf8]{inputenc}
\usepackage{enumitem}

\usepackage{tikz}
\usetikzlibrary{automata,positioning,chains,shapes,arrows,patterns}
\usepackage{pgfplots}
\usetikzlibrary{plotmarks}
\newlength\fheight
\newlength\fwidth
\pgfplotsset{compat=newest}
\pgfplotsset{plot coordinates/math parser=false}

\usepackage{array}
\usepackage[font=scriptsize]{caption}
\usepackage[font=scriptsize]{subcaption}

\usepackage[top=1.5cm, bottom=2cm, right=1.6cm,left=1.6cm]{geometry}
\usepackage{indentfirst}

\usepackage{times}

\usepackage{breqn}
\usepackage{booktabs}
\usepackage{etoolbox}
\usepackage{placeins}
\newif\iftikz
\tikztrue
\begin{document}
\title{TCP in 5G mmWave Networks:\\Link Level Retransmissions and MP-TCP}

\author{Michele Polese$^*$, Rittwik Jana$^\dagger$, Michele Zorzi$^*$\\
 \small $^*$Department of Information Engineering, University of Padova, Italy \\
e-mail: \{polesemi, zorzi\}@dei.unipd.it\\
$^\dagger$AT\&T Labs-Research, Bedminster NJ, USA 
\\ e-mail: rjana@research.att.com}

\setlength{\belowcaptionskip}{-0.2cm}

\makeatletter
\patchcmd{\@maketitle}
  {\addvspace{0.5\baselineskip}\egroup}
  {\addvspace{0\baselineskip}\egroup}
  {}
  {}
\makeatother






\maketitle

\tikzstyle{startstop} = [rectangle, rounded corners, minimum width=2cm, minimum height=0.5cm,text centered, draw=black]
\tikzstyle{io} = [trapezium, trapezium left angle=70, trapezium right angle=110, minimum width=3cm, minimum height=1cm, text centered, draw=black]
\tikzstyle{process} = [rectangle, minimum width=2cm, minimum height=0.5cm, text centered, draw=black, alignb=center]
\tikzstyle{decision} = [ellipse, minimum width=2cm, minimum height=1cm, text centered, draw=black]
\tikzstyle{arrow} = [thick,<->,>=stealth]
\tikzstyle{line} = [thick,>=stealth]
\tikzstyle{darrow} = [thick,<->,>=stealth,dashed]
\tikzstyle{sarrow} = [thick,->,>=stealth]
\tikzstyle{larrow} = [line width=0.05mm,dashdotted,>=stealth]

\begin{abstract}
MmWave communications, one of the cornerstones of future 5G mobile networks, are characterized at the same time by a potential multi-gigabit capacity and by a very dynamic channel, sensitive to blockage, wide fluctuations in the received signal quality, and possibly also sudden link disruption. While the performance of physical and MAC layer schemes that address these issues has been thoroughly investigated in the literature, the complex interactions between mmWave links and transport layer protocols such as TCP are still relatively unexplored. This paper uses the ns--3 mmWave module, with its channel model based on real measurements in New York City, to analyze the performance of the Linux TCP/IP stack (i) with and without link-layer retransmissions, showing that they are fundamental to reach a high TCP throughput on mmWave links and (ii) with Multipath TCP (MP-TCP) over multiple LTE and mmWave links, illustrating which are the throughput-optimal combinations of secondary paths and congestion control algorithms in different conditions.
\end{abstract}
\begin{picture}(0,0)(0,-360)
\put(0,0){
\put(0,0){\footnotesize This paper was accepted for presentation at the 2017 IEEE Infocom} 
\put(0,-10){\footnotesize 5G \& Beyond Workshop, May 1, 2017, Atlanta, Georgia, USA.}}
\end{picture}

\section{Introduction}
MmWave communications are expected to play a major role in reaching the performance target of the next generation of mobile networks (5G)~\cite{boccardi2014five}. At mmWave frequencies (i.e., above 10 GHz), indeed, there is a high availability of contiguous bandwidth that can be allocated to cellular networks. However, mmWave communications also present challenges and issues that must be faced in order to make this technology market-ready. In fact, these frequencies suffer from high isotropic pathloss (compensated by massive MIMO and beamforming gains) and blockage by solid materials, as for example buildings, cars, and also the human body~\cite{rappaport1}.

These extreme propagation conditions demand a careful design of the PHY and MAC layers~\cite{zorzimac}, but also have an impact on the interplay with the higher layers of the protocol stack. In particular, the congestion control mechanisms of the Transmission Control Protocol (TCP) may suffer from long blockages that trigger a Retransmission Timeout (RTO), and may not timely track the channel state when the link from a User Equipment (UE) to the serving evolved Node Base (eNB) switches from a Line-of-Sight (LOS) to a Non-Line-of-Sight (NLOS) condition~\cite{mmNet}. Finally, TCP may take a long time to fill the huge amount of bandwidth available, and this penalizes short-lived TCP sessions such as those used for browsing or instant messaging applications.

In this paper, we systematically analyze the performance of the Linux kernel TCP/IP stack implementation over mmWave links using the mmWave module for the ns--3 simulator. 
First, we study the interaction with lower-layer retransmission protocols, in terms of both throughput and latency, proving that without these retransmissions TCP is only able to reach a fraction of the potential mmWave NLOS link throughput. Secondly, we show that multi-path transmissions improve the performance of the mmWave network, by using Multipath TCP (MP-TCP) with different congestion control algorithms over (i) an LTE and a mmWave link or (ii) two mmWave links with different carrier frequencies. Finally, we test the multi-path transmission of TCP ACKs only, measuring in which conditions sending the TCP ACK packets over an LTE link (and data over a mmWave one) improves the throughput in a mobility scenario.

The rest of the paper is organized as follows. Sec.~\ref{sec:setup} describes the simulation setup, while Sec.~\ref{sec:retx} illustrates the performance analysis involving the lower-layer retransmission. Sec~\ref{sec:mp} presents the results for MP-TCP in mmWave networks, and Sec.~\ref{sec:mpack} those for TCP ACK transmission on LTE links. Finally in Sec.~\ref{sec:concl} some conclusions are drawn, and future work is suggested.

\section{Simulation Setup}\label{sec:setup}
The simulations use the NYU ns--3 module~\cite{phy5g,mmWaveSim}, with the extensions developed in~\cite{simutoolspolese}, plugged to the TCP implementation of the Linux kernel (also including MP-TCP~\cite{mptcpImpl}) using a custom version of the Direct Code Execution (DCE) library~\cite{dce}. In this way it is possible to test the real Linux implementation of TCP and MP-TCP with the flexibility of a network simulator. This approach can be applied to any application layer software or transport protocol, provided they only use the subset of kernel methods also available in DCE. It is possible to plug different applications on top, and the following experiments use IPERF~\cite{iperf} and Linux wget, in order to test respectively the throughput of the end-to-end connection and the time it takes to download files of different sizes.

\begin{table}[b]
  \centering
  \begin{tabular}{@{}ll@{}}
  	\toprule
    Parameter & Value \\
    \midrule
    mmWave carrier frequency & 28 GHz, 73 GHz \\
    mmWave bandwidth & 1 GHz \\
    mmWave TX power & $30$ dBm \\
    LTE carrier frequency (DL) & 2.1 GHz \\
    LTE carrier frequency (UL) & 1.9 GHz \\
    LTE bandwidth & 20 MHz \\
    LTE downlink TX power & $30$ dBm \\
	LTE uplink TX power & $25$ dBm \\
  \bottomrule
  \end{tabular}
  \caption{Simulation parameters}
  \label{table:params}
\end{table}

The NYU ns--3 module~\cite{mmWaveSim} simulates an end-to-end cellular network, with a complete mobile stack (with custom TDD-based PHY and MAC layers, RRC layer, and legacy LTE RLC and PDCP layers) which is able to transmit packets from the UE to a remote host, or vice versa. The main feature of the NYU ns--3 module is the channel model for the 28 GHz and the 73 GHz carrier frequencies, based on real measurements~\cite{rappaport2}, which can either statistically simulate LOS-NLOS-outage transitions~\cite{phy5g} or rely on the ns--3 building module to track when a mobile terminal is in LOS or NLOS~\cite{mmWaveSim, poleseHo}. The main simulation parameters are those typically used in the performance analysis of mmWave networks~\cite{poleseHo}, and are summarized in Table~\ref{table:params}.

\section{Interaction With lower-layer Retransmission Protocols}
\label{sec:retx}

Current and future mobile networks deploy different retransmission mechanisms in order to prevent packet loss and increase the throughput at the mobile devices. When using mmWave links, these retransmission protocols become a key element in hiding the highly dynamic and consequently unstable behavior of the channel to the higher layer transport protocols. 

At the MAC layer, Hybrid Automatic Repeat reQuest (HARQ) is used. When the PHY layer at the receiver receives a packet, but detects the presence of some errors that prevent reliable decoding, it asks for a retransmission. The sender then transmits additional redundancy that helps retrieve the correct version of the packet~\cite{harq}. Moreover, in 3GPP networks (e.g., LTE), there is a layer on top of the MAC layer that may perform additional retransmissions, called Radio Link Control (RLC) layer, that will also likely be a part of the 5G protocol stack~\cite{ktRlc}. Since the number of retransmissions at the MAC layer is usually limited (typically only 3 attempts are performed), the RLC layer Acknowledged Mode (AM) offers another way of recovering lost packets. Thanks to periodic reports from the receiver, the RLC AM sender knows which packets are missing and can retransmit them. The number of attempts that RLC AM can perform is also limited, and, if some packets are still missing, a Radio Link Failure is declared. RLC Unacknowledged Mode instead does not perform any retransmission in addition to those of the HARQ at the MAC layer. These retransmission mechanisms operate based on information related to the link and with a greater timeliness with respect to TCP, which instead uses packet losses to detect congestion and operates on the larger timescale of retransmission time-outs (RTOs), of the order of a second.

In order to test the effectiveness of coupling TCP with lower-layer retransmission mechanisms, we performed some simulations using the framework described in Sec.~\ref{sec:setup}, where we considered an uplink connection from a User Equipment (UE) placed at different distances from an evolved Node Base (eNB). We use IPERF on top of the Linux implementation of TCP CUBIC, with the statistical channel model~\cite{phy5g}, and perform Montecarlo simulations for each distance $d\in\{50, 75, 100, 150\}$~m. 

\begin{figure}
	\centering
\iftikz
	\setlength\fwidth{0.8\columnwidth}
  	\setlength\fheight{0.5\columnwidth}
  	\input{figures/dist-harq.tex}
\else
	\includegraphics[width=0.8\columnwidth]{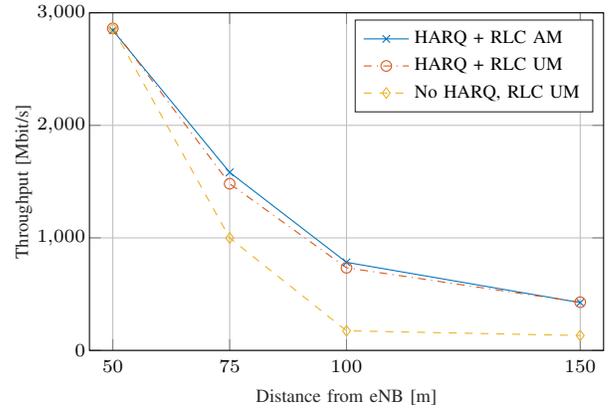}
\fi
	\caption{Throughput of TCP CUBIC with and without the different retransmission mechanisms of the mmWave protocol stack.}
	\label{fig:retxTh}
\end{figure}

\begin{figure}
	\centering
\iftikz
	\setlength\fwidth{0.8\columnwidth}
  	\setlength\fheight{0.5\columnwidth}
  	\input{figures/harqam.tex}
\else
	\includegraphics[width=0.8\columnwidth]{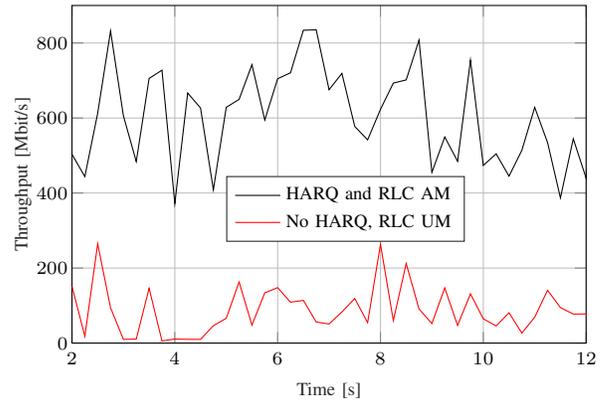}
\fi
	\caption{Example of TCP CUBIC flow with and without HARQ and RLC AM retransmissions. The UE is at 150 m from the eNB.}
	\label{fig:harqam}
\end{figure}

RLC AM introduces additional redundancy in order to perform the retransmissions, but, when the distance between the eNB and the UE is equal to $d=50$~m and the UE is in LOS with very high probability, these retransmissions are not actually needed, because of the low packet error rate of the channel. Therefore, as also shown in Fig.~\ref{fig:retxTh}, the throughput is lower when RLC AM is used (though by only a minimal amount). As the distance increases, instead, the performance of TCP without HARQ and without RLC AM collapses, because the TCP congestion control algorithm sees a very lossy link and triggers congestion avoidance mechanisms or, worse, a RTO. Fig.~\ref{fig:harqam} compares the traces at the PDCP layer of the throughput of a simulation over time, for $d=150$~m. It can be seen that the lack of HARQ and RLC AM places the whole burden of retransmissions on TCP, which does not manage to reach the high throughput allowed by the available bandwidth. 

If instead we compare the performance of HARQ with RLC UM and that of HARQ with RLC AM, it can be seen from Fig.~\ref{fig:retxTh} that the additional retransmissions given by RLC AM increase the throughput by 100 Mbit/s at $d=75$~m and 50 Mbit/s at $d=100$~m. For $d=150$~m, instead, RLC AM does not improve the performance of RLC UM, showing that at such distance even further transmission attempts fail to successfully deliver packets (for example, because of extended outage events).

\begin{figure}
	\centering	
\iftikz
	\setlength\fwidth{0.8\columnwidth}
  	\setlength\fheight{0.5\columnwidth}
  	\input{figures/latency-th.tex}
\else
	\includegraphics[width=0.8\columnwidth]{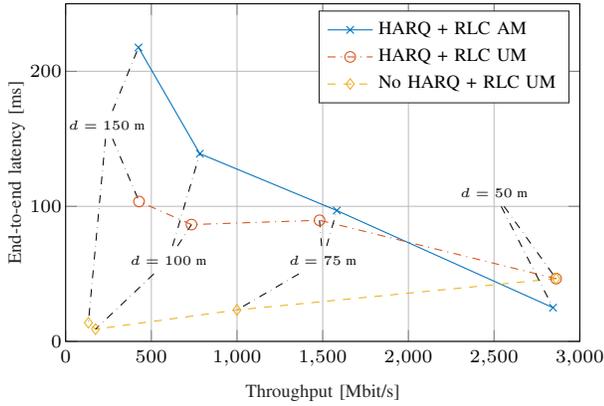}
\fi
	\caption{Latency throughput tradeoff for TCP CUBIC, with and without the different retransmission mechanisms of the mmWave protocol stack.}
	\label{fig:retxLat}
\end{figure}

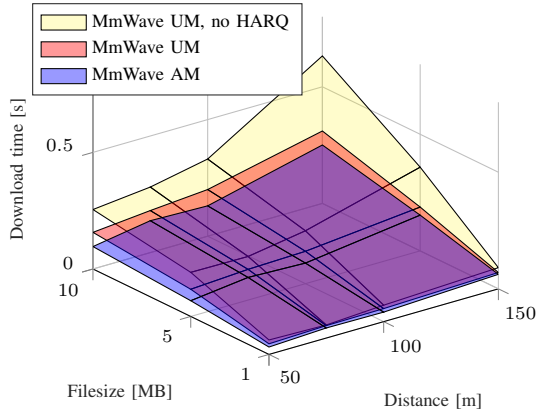
\begin{figure}
	\centering
	\setlength\fwidth{0.6\columnwidth}
  	\setlength\fheight{0.5\columnwidth}
  	\input{figures/dist-wget-harq.tex}
  	\caption{Download time as a function of the file size and of the distance, for TCP CUBIC with and without lower-layer retransmissions.}
  	\label{fig:wget-harq}
\end{figure}

RLC AM at large distances instead increases the latency of successfully received packets, as shown in Fig.~\ref{fig:retxLat}, because of retransmissions and additional segmentation that may introduce Head of Line (HoL) blocking delays. The smallest latency is achieved without HARQ and with RLC UM, because no retransmissions are performed, but this option is not able to deliver a high TCP throughput in general.

Fig.~\ref{fig:wget-harq} shows the download time for a file of different sizes (from 1 MB to 10 MB) using wget (the file is hosted in the UE and retrieved by the remote server, in order to be consistent with the previous uplink simulations). The results show that lower-layer retransmission mechanisms help decrease the download time, and that the performance gain increases as the distance and the file size increase. Moreover, the difference between the download times with RLC AM and with RLC UM (no retransmissions) is more noticeable than that between the throughput values of Fig.~\ref{fig:retxTh}, showing that for short-lived TCP sessions it is important to perform retransmissions as fast as possible, i.e., at a layer as close to the radio link as possible.

These results are well known when applied to traditional LTE networks~\cite{nekovee2016lte}, but these are the first simulations that show how much TCP depends on lower-layer retransmissions in mmWave networks, using the real Linux TCP/IP implementation. They show that, also in mmWave networks, the support of lower-layer retransmission mechanisms is fundamental for reaching a high TCP throughput even at large distances between transmitter and receiver, at the price of additional latency. In particular, in the simulated scenario the most effective retransmission scheme is HARQ at the MAC layer, since it provides the greatest throughput gain, but also the acknowledged mode of the RLC layer helps improve the performance of the mmWave link by reducing the download time for short-lived TCP sessions.

\section{Multipath TCP}
\label{sec:mp}
Multipath TCP (MP-TCP) has been proposed as a way of allowing vertical and seamless handovers between cellular networks and Wi-Fi hotspots and is currently under discussion for standardization at the IETF. It may also be used to provide path diversity in mmWave cellular networks. The three main design goals of MP-TCP are~\cite{rfc6356}:
\begin{enumerate}
	\item\label{obj:atleast} Improve throughput: an MP-TCP flow should perform at least as well as a traditional single path TCP (SP-TCP) flow on the best path available.
	\item\label{obj:noharm} On shared links, MP-TCP should not get more resources than standard TCP flows.
	\item\label{obj:lc} MP-TCP should prefer less congested paths, subject to the previous two conditions.
\end{enumerate}

There are three RFCs that describe MP-TCP~\cite{rfc6824,rfc6356,rfc6182}.
They discuss the signaling and setup procedures~\cite{rfc6824}, the architectural choices for the deployment of MP-TCP~\cite{rfc6182}, and a congestion control (CC) algorithm~\cite{rfc6356}. Finally the document in~\cite{app} discusses the impact on the application layer.

There are several studies that propose coupled congestion control algorithms for MP-TCP connections. By coupling over the different subflows, the authors of~\cite{rfc6356} claim that it is possible to reach goals~\ref{obj:noharm} and~\ref{obj:lc} above. In particular they propose a first coupled CC, that is however criticized in~\cite{olia} and in~\cite{balia}, because it (i) transmits too much traffic on congested paths and (ii) is unfriendly with respect to SP-TCP. Therefore two more coupled CC were proposed:
\begin{itemize}
	\item In~\cite{olia} the Opportunistic Linked Increases Algorithm (OLIA) is designed to overcome these two issues, but presents non-responsiveness problems with respect to congestion changes in the subflows;
	\item In~\cite{balia} the Balanced Linked Adaptation algorithm (BALIA) addresses both the problems of the original CC and those of OLIA. In particular, the parameters of the protocol are derived through a theoretical analysis of the performance of multipath congestion control algorithms.
\end{itemize}

However, these schemes are based on the legacy design of Reno and New Reno congestion control algorithms (Additive Increase - Multiplicative Decrease, AIMD), which are shown to suffer from the highly dynamic behavior of mmWave links more than the newer TCP CUBIC congestion control algorithm~\cite{mmNet}.

MP-TCP could be used as an end-to-end solution for multi-connectivity, i.e., next generation mobile devices may connect both to an LTE and to a mmWave eNB, or to two or more mmWave eNBs with no need for coordination at the lower layers. However, there are some issues with its performance in mmWave networks, as we will show in the following paragraphs. In this performance evaluation campaign we used the real Linux implementation of MP-TCP (v0.90), which includes several CC algorithms, namely the original coupled CC, OLIA, BALIA, uncoupled (with any desired TCP flavor, e.g., CUBIC), and others. We co-deploy an LTE eNB and a mmWave eNB, or a mmWave eNB capable of transmissions at different frequencies (28 and 73 GHz, with the same bandwidth and the maximum number of antennas available in the ns--3 NYU simulator), and vary the distance of the multi-connected UE from the eNBs, using the statistical channel model. The remote host is a multi-homed server, supporting MP-TCP connections. The UE uses IPERF, and starts the connection on the 28 GHz mmWave link. Then another subflow is added on the LTE link, or on the 73 GHz mmWave link.

\begin{figure}
	\centering
\iftikz
	\setlength\fwidth{0.85\columnwidth}
  	\setlength\fheight{0.5\columnwidth}
  	\input{figures/dist-mptcp.tex}
\else
	\includegraphics[width=\columnwidth]{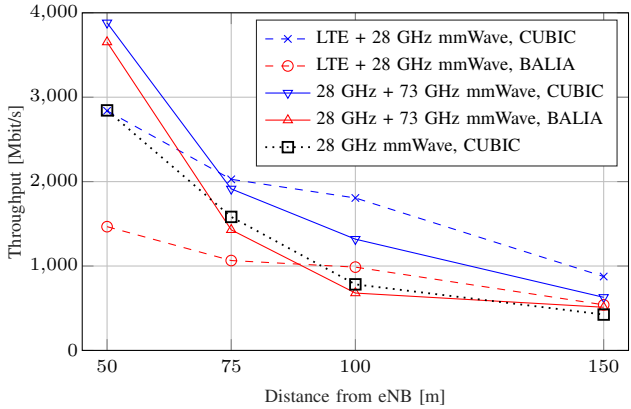}
\fi
	\caption{MP-TCP throughput for different distances $d$ and different MP-TCP options. The black dotted line shows the performance of a SP-TCP connection with TCP CUBIC, as a reference.}
	\label{fig:mptcp}
\end{figure}

Fig.~\ref{fig:mptcp} shows the performance in terms of throughput of different MP-TCP congestion control algorithms over different connections, with respect to the baseline of a SP-TCP connection with TCP CUBIC. The dashed lines represent a scenario with paths on LTE and on mmWave (28 GHz), while the solid ones refer to paths on mmWave links with 28 GHz and 73 GHz as carrier frequencies. 

\textbf{LTE as mmWave secondary path:} When the UE is close to the eNB and has a LOS link most of the time on both the 28 and the 73 GHz connections (e.g., for $d=50$~m), then the solution with multipath TCP on mmWave-only links outperforms SP-TCP, with a gain that ranges from 800 Mbit/s (28\%) to 1 Gbit/s (36\%). Instead, due to the limit of the LTE uplink, the performance of a multipath on LTE and mmWave is close to that of SP-TCP (when CUBIC is used, because BALIA has much worse performance, as will be discussed later).

\begin{figure}[t]
  \centering
\iftikz
  \setlength\fwidth{0.75\columnwidth}
    \setlength\fheight{0.49\columnwidth}
    \input{figures/dist-cubic-ratio-lte.tex}
\else
  \includegraphics[width=0.8\columnwidth]{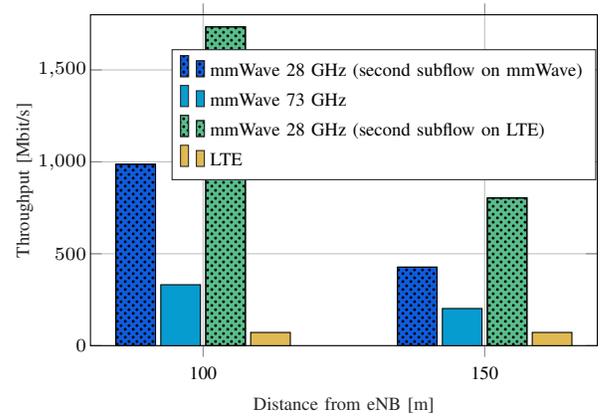}
\fi
  \caption{Contribution of the two subflows as a function of the distance $d$, for MP-TCP with CUBIC CC.}
  \label{fig:ratio}
\end{figure}

However, it can be seen from Fig.~\ref{fig:mptcp} that MP-TCP with LTE and mmWave links performs better than with only mmWave connections for $d\ge100$~m, and with the CUBIC uncoupled CC algorithm also for $d=75$~m. Indeed, the 73 GHz link offers a potentially larger throughput than an LTE uplink connection, but it has a lossy behavior that penalizes the overall throughput, except for small distances. In particular, for $d=150$~m, MP-TCP with LTE and 28 GHz mmWave offers a gain of more than 450 Mbit/s (i.e., 100\%) with respect to the SP-TCP (i.e., more than the LTE uplink throughput), showing that the presence of the secondary and reliable LTE path improves the throughput on the mmWave link. This can be seen also in Fig.~\ref{fig:ratio}, where we plot the contribution of the two subflows of MP-TCP connections at $d\in\{100, 150\}$~m when the second subflow is LTE or mmWave. It can be seen that the contribution given by the reliable LTE uplink subflow is smaller than that of the 73 GHz mmWave subflow, but the primary 28 GHz mmWave subflow reaches a higher throughput when coupled with the LTE secondary subflow.

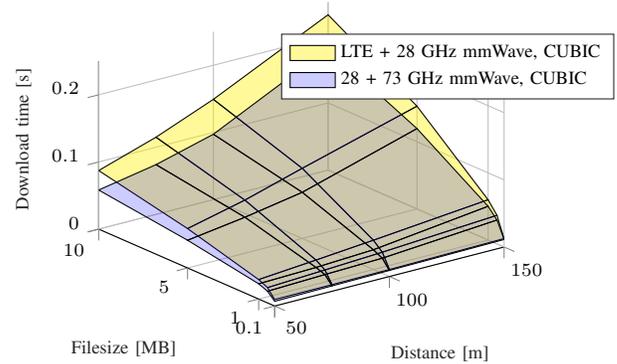
\begin{figure}[t]
  \centering
  \setlength\fwidth{0.6\columnwidth}
    \setlength\fheight{0.45\columnwidth}
    \input{figures/wget-mptcp-cubic.tex}
    \caption{Download time as a function of the file size and of the distance, for MP-TCP with CUBIC CC and secondary subflow on LTE or mmWave at 73 GHz.}
    \label{fig:wget-mptcp}
\end{figure}

\begin{figure}
  \centering
\iftikz
  \begin{subfigure}[t]{\columnwidth}
    \centering
    \setlength\fwidth{0.75\columnwidth}
      \setlength\fheight{0.83\columnwidth}
      \input{figures/olia-time.tex}
    \caption{OLIA CC}
  \end{subfigure}
  \vspace{1pt}
  \begin{subfigure}[t]{\columnwidth}
    \centering
    \setlength\fwidth{0.75\columnwidth}
      \setlength\fheight{0.83\columnwidth}
      \input{figures/balia-time.tex}
    \caption{BALIA CC}
  \end{subfigure}
  \vspace{1pt}
  \begin{subfigure}[t]{\columnwidth}
    \centering
    \setlength\fwidth{0.75\columnwidth}
      \setlength\fheight{0.83\columnwidth}
      \input{figures/cubic-time.tex}
    \caption{CUBIC uncoupled CC}
  \end{subfigure}
  
\else
  \includegraphics[width=0.9\columnwidth]{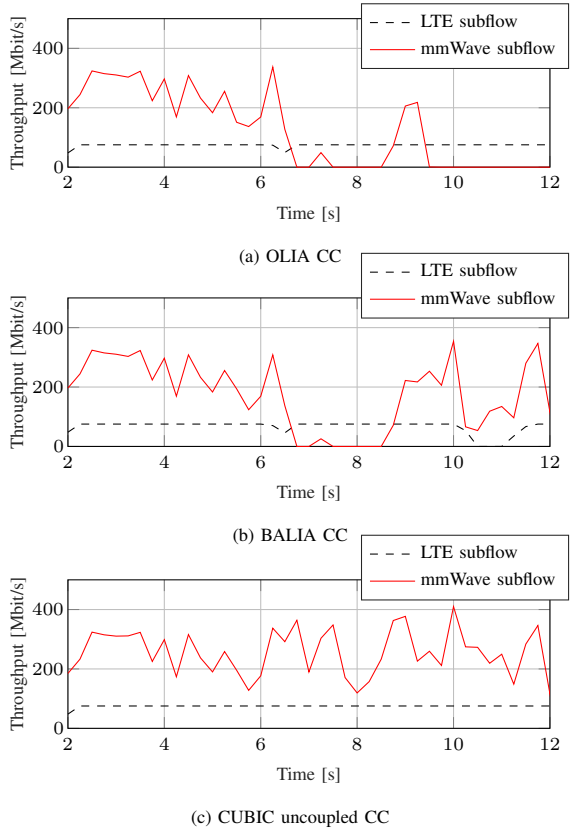}
\fi
  \caption{Throughput over time for the two subflows, for different MP-TCP CC algorithms.}
  \label{fig:thInTime}
\end{figure} 

For short-lived TCP sessions, instead, using a secondary subflow on mmWave links improves the system performance. This can be seen in Fig.~\ref{fig:wget-mptcp}, which shows the download time of a file using wget with the same setup described in Sec.~\ref{sec:retx}. However, the performance gain, especially for smaller files, is minimal, showing that the LTE link makes up for its smaller capacity with a higher reliability that benefits the performance of TCP.

\textbf{Coupled vs uncoupled CC}: Another important observation is that MP-TCP with the BALIA CC algorithm fails to meet target~\ref{obj:atleast}, since in many cases its throughput is lower than that of SP-TCP, as shown in Fig.~\ref{fig:mptcp}. The most striking cases are those with MP-TCP on LTE and mmWave, and $d\in\{50, 75\}$~m. Here the congestion control algorithm sees the losses on the 28 GHz mmWave link as congestion, and, according to design goal~\ref{obj:lc}, it steers the whole traffic to the LTE subflow, degrading the performance of the end-to-end connection. Instead, the uncoupled congestion control algorithm is not affected by this issue, since each path behaves independently. However, in this case design goal~\ref{obj:noharm} is not met. An example of this behavior is shown in Fig.~\ref{fig:thInTime}, where we compare the throughput over time of two different AIMD coupled CC algorithms (OLIA and BALIA) and of an uncoupled CC algorithm with CUBIC. It can be seen that at time $t=7$~s both OLIA and BALIA start using only the LTE connection, and that the throughput of the mmWave subflow goes to zero. A similar behavior for OLIA was observed in~\cite{balia}. 

\begin{figure}[t]
\setlength{\belowcaptionskip}{-0.2cm}
\begin{subfigure}[t]{\columnwidth}
  \centering
  \setlength\fwidth{0.6\columnwidth}
    \setlength\fheight{0.45\columnwidth}
    \input{figures/wget-mptcp-lte-balia-cubic-small.tex}
    \caption{Small file size.}
    \label{fig:wget-sf}
\end{subfigure}
\par\bigskip
\begin{subfigure}[b]{\columnwidth}
  \centering
  \setlength\fwidth{0.6\columnwidth}
    \setlength\fheight{0.45\columnwidth}
    \input{figures/wget-mptcp-lte-balia-cubic-large.tex}
    \caption{Large file size.}
    \label{fig:wget-lf}
\end{subfigure}
\caption{Download time as a function of the file size and of the distance, for MP-TCP with secondary subflow on LTE and CUBIC or BALIA CC.}
\label{fig:wget-balia-cubic}
\end{figure}
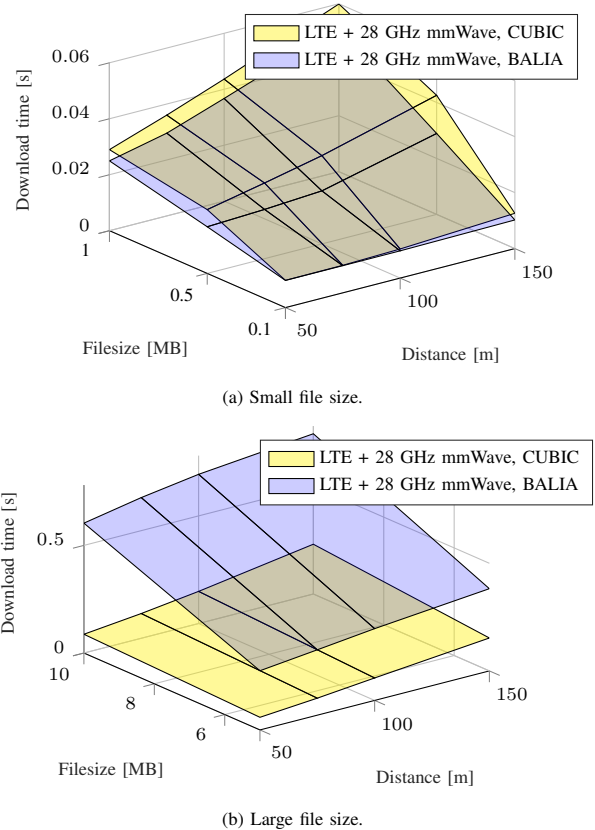

When considering short-lived TCP sessions and file download times, there are two different outcomes according to the file size. As shown in Fig.~\ref{fig:wget-sf}, when the file is smaller than 1 MB the BALIA coupled congestion control algorithm exhibits a slightly smaller download time than the CUBIC uncoupled CC. Instead, when the file is larger than 5 MB, as in Fig.~\ref{fig:wget-lf}, the MP-TCP solution with CUBIC as CC mechanism manages to download the file in less than a fifth of the time required by MP-TCP with BALIA. This behavior can be explained by considering the shape of the window growth function of CUBIC, which recalls a cubic function, i.e., flat at the beginning and then rapidly increasing. 

The main conclusions from this performance analysis of MP-TCP for mmWave networks are that at larger distances and for long-lived TCP sessions it is preferable to use a more stable LTE-like link, and that the deployment of MP-TCP coupled congestion control algorithms on mmWave links is not able to satisfy the original design goals of~\cite{rfc6356}. A possible improvement of MP-TCP CC algorithms should adapt the TCP CUBIC scheme to a coupled scenario, so that the reactiveness and stability of TCP CUBIC enhance the performance of the transport protocol while not harming other legacy TCP flows.

\section{Multi-connected UE with ACKs on LTE and data on mmWave}
\label{sec:mpack}

\begin{figure}[t!]
\centering
\begin{tikzpicture}[font=\sffamily, scale=0.54, every node/.style={scale=0.54}]
  \centering

    \node[anchor=south west,inner sep=0] (image) at (0,0) {\includegraphics[width=0.7\textwidth]{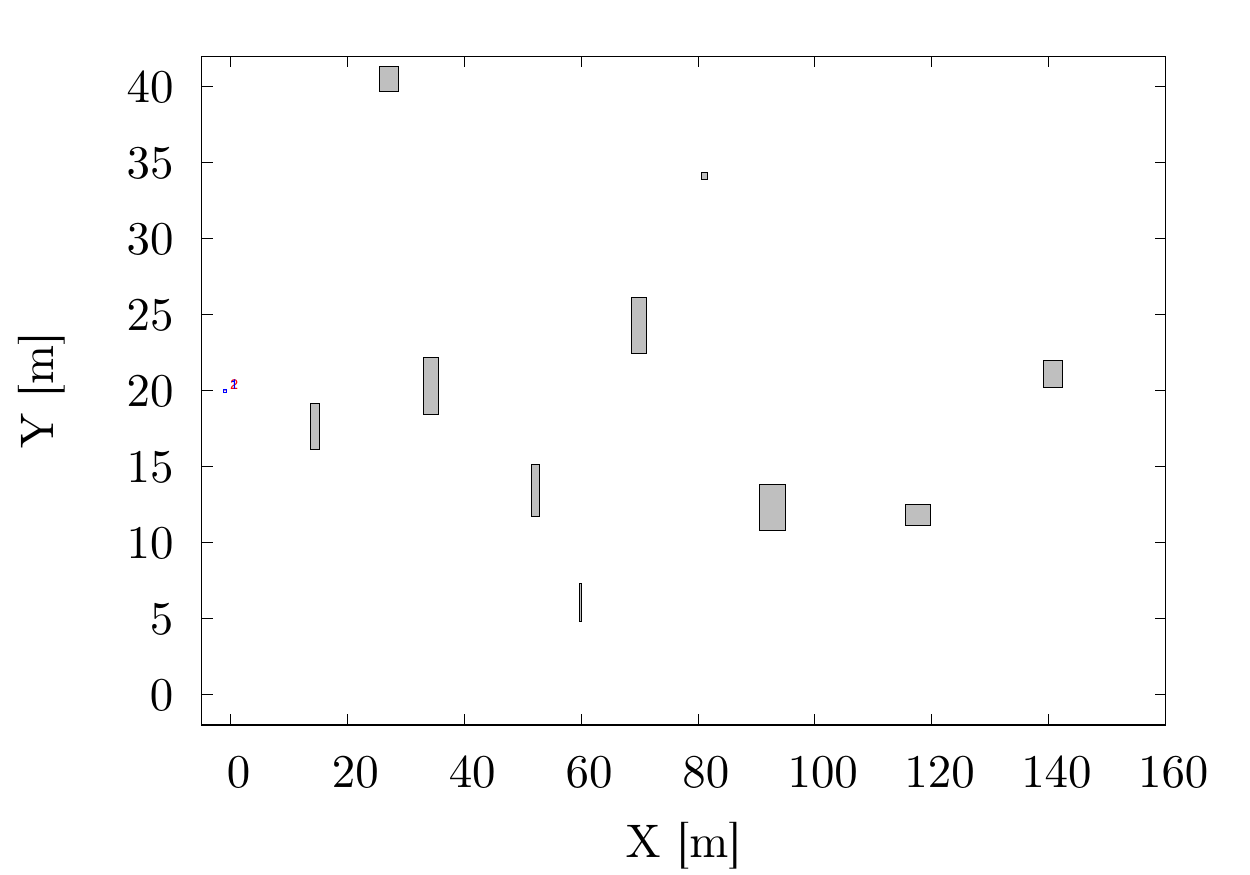}};
    \begin{scope}[x={(image.south east)},y={(image.north west)}]
        \filldraw[red,ultra thick] (0.18,0.56) circle (2pt);
        \node[anchor=south] at (0.19,0.575) (mm1label) {eNB};
        \draw[sarrow] (0.884, 0.21) -- (0.884, 0.9);
        \node[anchor=east] at (0.884, 0.71) (arrowLabel) {UE path at speed $v$};

        \filldraw[blue,ultra thick] (0.884, 0.207) circle (2pt);
        \node[anchor=east] at (0.88, 0.207) (ltelabel) {UE};
    \end{scope}   
\end{tikzpicture}
\caption{Random realization of the simulation scenario. The grey rectangles are $10$ randomly deployed non-overlapping obstacles (e.g., cars, buildings, people, trees).}
\label{fig:map}
\end{figure}
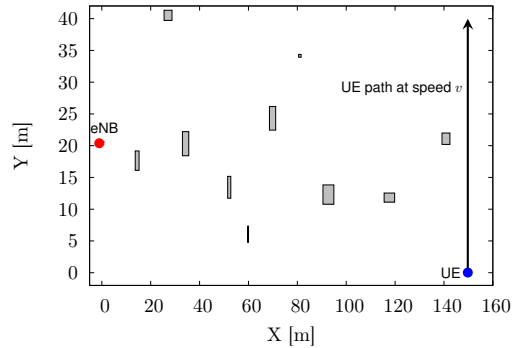

In this section, we test the performance of another kind of multipath deployment. The UE receives downlink data from the eNB on a mmWave connection, and sends the TCP ACKs either on LTE or on the same mmWave link. We consider a mobility scenario, with the eNB at coordinates (-1, 20)~m and the UE moving from (151, 0)~m to (151, 40)~m at speed $s\in\{2,5\}$~m/s. 10 small obstacles are deployed randomly in the area between the eNB and the UE, and the simulations exploit an ns--3 channel model that uses real traces to model the LOS to NLOS (or vice versa) transitions~\cite{poleseHo}. An example of random deployment is shown in Fig.~\ref{fig:map}.

\begin{figure}
\setlength{\belowcaptionskip}{-0.4cm}
	\centering
\iftikz
	\setlength\fwidth{0.8\columnwidth}
  	\setlength\fheight{0.5\columnwidth}
  	\input{figures/throughput-buffer-speed.tex}
\else
	\includegraphics[width=0.8\columnwidth]{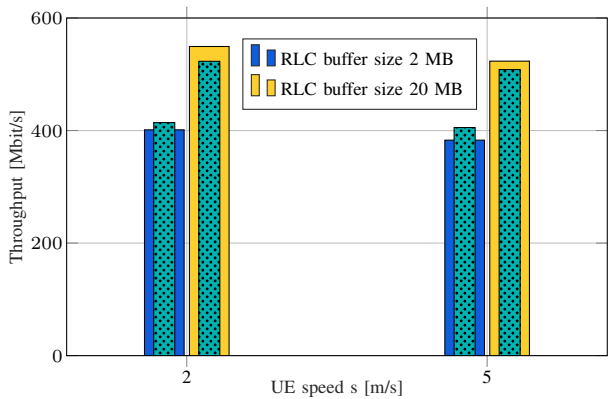}
\fi
	\caption{TCP throughput for different ACK reporting links. The narrow dotted bars refer to the setup with ACKs on mmWave links, while the large ones to that with ACKs on the LTE link.}
	\label{fig:tcpAck}
\end{figure}

The results of these simulations are shown in Fig.~\ref{fig:tcpAck} for different RLC buffer sizes (2, 20 MB) and UE speed $s$ (2, 5 m/s). It can be seen that with a larger buffer the throughput is slightly higher when sending the ACKs on the LTE connection, while with the smaller one it is preferable to use the mmWave connection, but the throughput is 100 Mbit/s lower. Indeed, when the buffer is small there is a need for more timely updates of the TCP congestion window, since there are more chances to cause buffer overflow and lose packets. If LTE is used, the latency of the ACKs increases, therefore the timeliness of congestion control is reduced. However, the difference in throughput between the two solutions is small. Instead, with a larger buffer
it is possible to queue more packets, and the transport layer is less sensitive to the latency in reporting the ACKs. In this case it is better to receive the ACKs on a more reliable LTE connection. However, notice that when the ACKs are sent on LTE the RTT increases, thus it takes more time to fill the capacity of the LOS link. This explains why also for the large buffer case the difference in performance between the system with ACKs on LTE and that with ACKs on mmWave is minimal. The main difference is thus on the choice of the buffer size: an undersized buffer degrades the throughput up to 27\%.

\section{Conclusions}
\label{sec:concl}
The large bandwidth available at mmWave frequencies could allow a link capacity of the order of gigabits per second. However, the interaction with legacy transport protocols could prevent the full exploitation of the rate potentially available in mmWave communications. In this paper, we presented the first comprehensive performance evaluation of the interaction of Single Path and Multi Path TCP with mmWave links, with and without lower-layer retransmission mechanisms, using (i) a simulator with a channel model based on real measurements and a complete 3GPP-like cellular stack, and (ii) the actual TCP implementation of Linux. We firstly remarked that for mmWave it is very important to mask the channel losses to the higher TCP layer with link retransmission mechanisms, otherwise it is not possible to reach high throughput. Secondly, we studied the behavior of MP-TCP on 28 GHz mmWave links with LTE or 73 GHz mmWave links as secondary subflows. We showed that when the mmWave link has a high probability of being in NLOS state, a secondary LTE subflow improves the throughput performance of long-lived TCP sessions more than a mmWave subflow, and that the design goals of MP-TCP may not be met with mmWave links. Finally, we evaluated whether or not using LTE as the uplink connection for TCP ACKs helps improve the throughput, and showed that there is not a clear gain, because of the additional latency introduced by the LTE radio link. 

As part of our future work, we will study how it is possible to exploit connections over multiple paths and lower-layer retransmission mechanisms to reach a high throughput on mmWave links, while trying to reduce the additional delay introduced by retransmissions to meet the 1 ms latency 5G design goal. For example, we will extend our study to account for multiple mmWave eNBs deployed in different locations, so that the UE can use more than two MP-TCP subflows.

\bibliographystyle{IEEEtran}
\bibliography{main.bib}{}  

\end{document}

%% file: figures/dist-harq.tex
%
%
\definecolor{mycolor1}{rgb}{0.00000,0.44700,0.74100}%
\definecolor{mycolor2}{rgb}{0.85000,0.32500,0.09800}%
\definecolor{mycolor3}{rgb}{0.92900,0.69400,0.12500}%
\begin{tikzpicture}
\pgfplotsset{every tick label/.append style={font=\scriptsize}}

\begin{axis}[%
width=0.951\fwidth,
height=\fheight,
at={(0\fwidth,0\fheight)},
scale only axis,
xmin=45,
xmax=155,
xtick={ 50,  75, 100, 150},
xlabel style={font=\scriptsize\color{white!15!black}},
xlabel={Distance from eNB [m]},
ymin=0,
ymax=3000,
ylabel style={font=\scriptsize\color{white!15!black}},
ylabel={Throughput [Mbit/s]},
axis background/.style={fill=white},
xmajorgrids,
ymajorgrids,
ylabel shift = -5 pt,
yticklabel shift = -2 pt,
legend style={font=\scriptsize,legend cell align=left, align=left, draw=white!15!black}
]
\addplot [color=mycolor1, mark=x, mark options={solid, mycolor1}]
  table[row sep=crcr]{%
50	2843.81612863636\\
75	1582.09386244156\\
100	782.847855418182\\
150	424.976978755\\
};
\addlegendentry{HARQ + RLC AM}

\addplot [color=mycolor2, dashdotted, mark=o, mark options={solid, mycolor2}]
  table[row sep=crcr]{%
50	2861.71746154545\\
75	1481.38169514545\\
100	734.278350909091\\
150	428.50228404091\\
};
\addlegendentry{HARQ + RLC UM}

\addplot [color=mycolor3, dashed, mark=diamond, mark options={solid, mycolor3}]
  table[row sep=crcr]{%
50	2861.71746154545\\
75	999.3980034\\
100	174.251153195652\\
150	132.522453101604\\
};
\addlegendentry{No HARQ, RLC UM}

\end{axis}
\end{tikzpicture}%

%% file: figures/harqam.tex
%
%
\begin{tikzpicture}
\pgfplotsset{every tick label/.append style={font=\scriptsize}}

\begin{axis}[%
width=0.951\fwidth,
height=\fheight,
at={(0\fwidth,0\fheight)},
scale only axis,
xmin=2,
xmax=12,
xlabel style={font=\scriptsize\color{white!15!black}},
xlabel={Time [s]},
ymin=0,
ymax=900,
ylabel style={font=\scriptsize\color{white!15!black}},
ylabel={Throughput [Mbit/s]},
axis background/.style={fill=white},
xmajorgrids,
ymajorgrids,
ylabel shift = -5 pt,
yticklabel shift = -2 pt,
legend style={font=\scriptsize,at={(0.3,0.3)}, anchor=south west, legend cell align=left, align=left, draw=white!15!black}
]

\addplot [color=black]
  table[row sep=crcr]{%
2	503.238208\\
2.25	443.87104\\
2.5	612.912128\\
2.75	831.795584\\
3	605.942848\\
3.25	483.908352\\
3.5	705.531456\\
3.75	727.496704\\
4	371.871168\\
4.25	666.64768\\
4.5	626.75456\\
4.75	408.736256\\
5	628.4368\\
5.25	649.58496\\
5.5	742.060096\\
5.75	594.695872\\
6	704.762432\\
6.25	720.527424\\
6.5	834.006528\\
6.75	835.063936\\
7	675.395328\\
7.25	718.941312\\
7.5	577.392832\\
7.75	541.777408\\
8	623.101696\\
8.25	693.08288\\
8.5	701.590208\\
8.75	807.667456\\
9	455.069952\\
9.25	549.37152\\
9.5	484.72544\\
9.75	754.172224\\
10	473.574592\\
10.25	504.527808\\
10.5	444.976512\\
10.75	513.948352\\
11	628.388736\\
11.25	534.519744\\
11.5	387.011328\\
11.75	544.613184\\
12	439.641408\\
12.25	739.080128\\
12.5	499.8656\\
};
\addlegendentry{HARQ and RLC AM}

\addplot [color=red]
  table[row sep=crcr]{%
2	151.3136\\
2.25	18.024\\
2.5	264.544256\\
2.75	94.013184\\
3	9.901184\\
3.25	10.718272\\
3.5	146.018432\\
3.75	5.671552\\
4	10.8144\\
4.25	9.949248\\
4.5	9.901184\\
4.75	46.333696\\
5	66.088\\
5.25	162.744704\\
5.5	48.015936\\
5.75	134.002432\\
6	147.7968\\
6.25	108.913024\\
6.5	113.91168\\
6.75	56.186816\\
7	50.851712\\
7.25	82.9104\\
7.5	118.814208\\
7.75	54.408448\\
8	263.198464\\
8.25	59.935808\\
8.5	212.010304\\
8.75	90.84096\\
9	52.005248\\
9.25	147.07584\\
9.5	47.150784\\
9.75	131.262784\\
10	64.742208\\
10.25	45.612736\\
10.5	80.795584\\
10.75	26.67552\\
11	69.019904\\
11.25	140.779456\\
11.5	94.638016\\
11.75	77.286912\\
12	77.719488\\
12.25	14.082752\\
12.5	69.740864\\
};
\addlegendentry{No HARQ, RLC UM}

\end{axis}
\end{tikzpicture}%

%% file: figures/latency-th.tex
%
%
\definecolor{mycolor1}{rgb}{0.00000,0.44700,0.74100}%
\definecolor{mycolor2}{rgb}{0.85000,0.32500,0.09800}%
\definecolor{mycolor3}{rgb}{0.92900,0.69400,0.12500}%
\begin{tikzpicture}
\pgfplotsset{every tick label/.append style={font=\scriptsize}}
\begin{axis}[%
width=0.951\fwidth,
height=\fheight,
at={(0\fwidth,0\fheight)},
scale only axis,
xmin=0,
xmax=3000,
xlabel style={font=\scriptsize\color{white!15!black}},
xlabel={Throughput [Mbit/s]},
xmajorgrids,
ymin=0,
ymax=250,
ylabel style={font=\scriptsize\color{white!15!black}},
ylabel={End-to-end latency [ms]},
ymajorgrids,
ylabel shift = -5 pt,
yticklabel shift = -2 pt,
axis background/.style={fill=white},
legend style={font=\scriptsize,legend cell align=left,align=left,draw=white!15!black},
]
\addplot [color=mycolor1,solid,mark=x,mark options={solid}]
  table[row sep=crcr]{%
2843.81612863636	24.9606146341463\\
1582.09386244156	96.9176856728057\\
782.847855418182	139.00325766182\\
424.976978755	217.628436884192\\
};
\addlegendentry{HARQ + RLC AM};

\addplot [color=mycolor2,dashdotted,mark=o,mark options={solid}]
  table[row sep=crcr]{%
2861.71746154545	46.4644734146342\\
1481.38169514545	89.6936816648798\\
734.278350909091	86.5063698870432\\
428.50228404091	103.584042658361\\
};
\addlegendentry{HARQ + RLC UM};

\addplot [color=mycolor3,dashed,mark=diamond,mark options={solid}]
  table[row sep=crcr]{%
2861.71746154545	46.4644734146342\\
999.3980034	23.2405540673365\\
174.251153195652	9.04771981860465\\
132.522453101604	13.8067618195532\\
};
\addlegendentry{No HARQ + RLC UM};
\node[anchor=south] at (240, 150) (d150) {\tiny{$d=150 $ m}};
\node[anchor=south] at (600, 50) (d100) {\tiny{$d=100 $ m}};
\node[anchor=south] at (1500, 50) (d75) {\tiny{$d=75 $ m}};
\node[anchor=south] at (2500, 100) (d50) {\tiny{$d=50 $ m}};
\draw[larrow] (d150) -- (132.522453101604,	13.8067618195532);
\draw[larrow] (d150) -- (428.50228404091,	103.584042658361);
\draw[larrow] (d150) -- (424.976978755,	217.628436884192);
\draw[larrow] (d100) -- (174.251153195652,	9.04771981860465);
\draw[larrow] (d100) -- (734.278350909091,	86.5063698870432);
\draw[larrow] (d100) -- (782.847855418182,	139.00325766182);
\draw[larrow] (d75) -- (999.3980034,	23.2405540673365);
\draw[larrow] (d75) -- (1481.38169514545,	89.6936816648798);
\draw[larrow] (d75) -- (1582.09386244156,	96.9176856728057);
\draw[larrow] (d50) -- (2861.71746154545,	46.4644734146342);
\draw[larrow] (d50) -- (2843.81612863636,	24.9606146341463);

\end{axis}
\end{tikzpicture}%

%% file: figures/dist-wget-harq.tex
%
%
\begin{tikzpicture}
\pgfplotsset{every tick label/.append style={font=\scriptsize}}

\begin{axis}[%
width=\fwidth,
height=\fheight,
at={(0.772in,0.516in)},
scale only axis,
xmin=50,
xmax=150,
tick align=outside,
xlabel style={font=\scriptsize\color{white!15!black}},
xlabel={Distance [m]},
ymin=1,
ymax=10,
ytick={1,5,10},
ylabel style={font=\scriptsize\color{white!15!black}},
ylabel={Filesize [MB]},
zmin=0,
zmax=0.8,
zlabel style={font=\scriptsize\color{white!15!black}},
zlabel={Download time [s]},
view={-37.5}{30},
axis background/.style={fill=white},
axis x line*=bottom,
axis y line*=left,
axis z line*=left,
xmajorgrids,
ymajorgrids,
zmajorgrids,
legend style={font=\scriptsize, at={(-0.15,0.767)}, anchor=south west, legend cell align=left, align=left, draw=white!15!black}
]

\addplot3[%
surf,
fill opacity=0.2, shader=flat corner, draw=black, z buffer=sort, colormap={mymap}{[1pt] rgb(0pt)=(0.2081,0.1663,0.5292); rgb(1pt)=(0.211624,0.189781,0.577676); rgb(2pt)=(0.212252,0.213771,0.626971); rgb(3pt)=(0.2081,0.2386,0.677086); rgb(4pt)=(0.195905,0.264457,0.7279); rgb(5pt)=(0.170729,0.291938,0.779248); rgb(6pt)=(0.125271,0.324243,0.830271); rgb(7pt)=(0.0591333,0.359833,0.868333); rgb(8pt)=(0.0116952,0.38751,0.881957); rgb(9pt)=(0.00595714,0.408614,0.882843); rgb(10pt)=(0.0165143,0.4266,0.878633); rgb(11pt)=(0.0328524,0.443043,0.871957); rgb(12pt)=(0.0498143,0.458571,0.864057); rgb(13pt)=(0.0629333,0.47369,0.855438); rgb(14pt)=(0.0722667,0.488667,0.8467); rgb(15pt)=(0.0779429,0.503986,0.838371); rgb(16pt)=(0.0793476,0.520024,0.831181); rgb(17pt)=(0.0749429,0.537543,0.826271); rgb(18pt)=(0.0640571,0.556986,0.823957); rgb(19pt)=(0.0487714,0.577224,0.822829); rgb(20pt)=(0.0343429,0.596581,0.819852); rgb(21pt)=(0.0265,0.6137,0.8135); rgb(22pt)=(0.0238905,0.628662,0.803762); rgb(23pt)=(0.0230905,0.641786,0.791267); rgb(24pt)=(0.0227714,0.653486,0.776757); rgb(25pt)=(0.0266619,0.664195,0.760719); rgb(26pt)=(0.0383714,0.674271,0.743552); rgb(27pt)=(0.0589714,0.683757,0.725386); rgb(28pt)=(0.0843,0.692833,0.706167); rgb(29pt)=(0.113295,0.7015,0.685857); rgb(30pt)=(0.145271,0.709757,0.664629); rgb(31pt)=(0.180133,0.717657,0.642433); rgb(32pt)=(0.217829,0.725043,0.619262); rgb(33pt)=(0.258643,0.731714,0.595429); rgb(34pt)=(0.302171,0.737605,0.571186); rgb(35pt)=(0.348167,0.742433,0.547267); rgb(36pt)=(0.395257,0.7459,0.524443); rgb(37pt)=(0.44201,0.748081,0.503314); rgb(38pt)=(0.487124,0.749062,0.483976); rgb(39pt)=(0.530029,0.749114,0.466114); rgb(40pt)=(0.570857,0.748519,0.44939); rgb(41pt)=(0.609852,0.747314,0.433686); rgb(42pt)=(0.6473,0.7456,0.4188); rgb(43pt)=(0.683419,0.743476,0.404433); rgb(44pt)=(0.71841,0.741133,0.390476); rgb(45pt)=(0.752486,0.7384,0.376814); rgb(46pt)=(0.785843,0.735567,0.363271); rgb(47pt)=(0.818505,0.732733,0.34979); rgb(48pt)=(0.850657,0.7299,0.336029); rgb(49pt)=(0.882433,0.727433,0.3217); rgb(50pt)=(0.913933,0.725786,0.306276); rgb(51pt)=(0.944957,0.726114,0.288643); rgb(52pt)=(0.973895,0.731395,0.266648); rgb(53pt)=(0.993771,0.745457,0.240348); rgb(54pt)=(0.999043,0.765314,0.216414); rgb(55pt)=(0.995533,0.786057,0.196652); rgb(56pt)=(0.988,0.8066,0.179367); rgb(57pt)=(0.978857,0.827143,0.163314); rgb(58pt)=(0.9697,0.848138,0.147452); rgb(59pt)=(0.962586,0.870514,0.1309); rgb(60pt)=(0.958871,0.8949,0.113243); rgb(61pt)=(0.959824,0.921833,0.0948381); rgb(62pt)=(0.9661,0.951443,0.0755333); rgb(63pt)=(0.9763,0.9831,0.0538)}, mesh/rows=4,area legend,fill=yellow]
table[row sep=crcr, point meta=\thisrow{c}] {%
x	y	z	c\\
50	1	0.0619471219367148	3\\
50	5	0.189101691325527	3\\
50	10	0.254979754607484	3\\
75	1	0.0560236643958408	3\\
75	5	0.186907503214809	3\\
75	10	0.281025179856115	3\\
100	1	0.0705420344111098	3\\
100	5	0.226771840179412	3\\
100	10	0.33097541494354	3\\
150	1	0.0907775236742581	3\\
150	5	0.358819830004283	3\\
150	10	0.632409968448997	3\\
};
\addlegendentry{MmWave UM, no HARQ}

\addplot3[%
surf,
fill opacity=0.4, shader=flat corner, draw=black, z buffer=sort, colormap={mymap}{[1pt] rgb(0pt)=(0.2081,0.1663,0.5292); rgb(1pt)=(0.211624,0.189781,0.577676); rgb(2pt)=(0.212252,0.213771,0.626971); rgb(3pt)=(0.2081,0.2386,0.677086); rgb(4pt)=(0.195905,0.264457,0.7279); rgb(5pt)=(0.170729,0.291938,0.779248); rgb(6pt)=(0.125271,0.324243,0.830271); rgb(7pt)=(0.0591333,0.359833,0.868333); rgb(8pt)=(0.0116952,0.38751,0.881957); rgb(9pt)=(0.00595714,0.408614,0.882843); rgb(10pt)=(0.0165143,0.4266,0.878633); rgb(11pt)=(0.0328524,0.443043,0.871957); rgb(12pt)=(0.0498143,0.458571,0.864057); rgb(13pt)=(0.0629333,0.47369,0.855438); rgb(14pt)=(0.0722667,0.488667,0.8467); rgb(15pt)=(0.0779429,0.503986,0.838371); rgb(16pt)=(0.0793476,0.520024,0.831181); rgb(17pt)=(0.0749429,0.537543,0.826271); rgb(18pt)=(0.0640571,0.556986,0.823957); rgb(19pt)=(0.0487714,0.577224,0.822829); rgb(20pt)=(0.0343429,0.596581,0.819852); rgb(21pt)=(0.0265,0.6137,0.8135); rgb(22pt)=(0.0238905,0.628662,0.803762); rgb(23pt)=(0.0230905,0.641786,0.791267); rgb(24pt)=(0.0227714,0.653486,0.776757); rgb(25pt)=(0.0266619,0.664195,0.760719); rgb(26pt)=(0.0383714,0.674271,0.743552); rgb(27pt)=(0.0589714,0.683757,0.725386); rgb(28pt)=(0.0843,0.692833,0.706167); rgb(29pt)=(0.113295,0.7015,0.685857); rgb(30pt)=(0.145271,0.709757,0.664629); rgb(31pt)=(0.180133,0.717657,0.642433); rgb(32pt)=(0.217829,0.725043,0.619262); rgb(33pt)=(0.258643,0.731714,0.595429); rgb(34pt)=(0.302171,0.737605,0.571186); rgb(35pt)=(0.348167,0.742433,0.547267); rgb(36pt)=(0.395257,0.7459,0.524443); rgb(37pt)=(0.44201,0.748081,0.503314); rgb(38pt)=(0.487124,0.749062,0.483976); rgb(39pt)=(0.530029,0.749114,0.466114); rgb(40pt)=(0.570857,0.748519,0.44939); rgb(41pt)=(0.609852,0.747314,0.433686); rgb(42pt)=(0.6473,0.7456,0.4188); rgb(43pt)=(0.683419,0.743476,0.404433); rgb(44pt)=(0.71841,0.741133,0.390476); rgb(45pt)=(0.752486,0.7384,0.376814); rgb(46pt)=(0.785843,0.735567,0.363271); rgb(47pt)=(0.818505,0.732733,0.34979); rgb(48pt)=(0.850657,0.7299,0.336029); rgb(49pt)=(0.882433,0.727433,0.3217); rgb(50pt)=(0.913933,0.725786,0.306276); rgb(51pt)=(0.944957,0.726114,0.288643); rgb(52pt)=(0.973895,0.731395,0.266648); rgb(53pt)=(0.993771,0.745457,0.240348); rgb(54pt)=(0.999043,0.765314,0.216414); rgb(55pt)=(0.995533,0.786057,0.196652); rgb(56pt)=(0.988,0.8066,0.179367); rgb(57pt)=(0.978857,0.827143,0.163314); rgb(58pt)=(0.9697,0.848138,0.147452); rgb(59pt)=(0.962586,0.870514,0.1309); rgb(60pt)=(0.958871,0.8949,0.113243); rgb(61pt)=(0.959824,0.921833,0.0948381); rgb(62pt)=(0.9661,0.951443,0.0755333); rgb(63pt)=(0.9763,0.9831,0.0538)}, mesh/rows=4,area legend,fill=red]
table[row sep=crcr, point meta=\thisrow{c}] {%
x	y	z	c\\
50	1	0.0454628114202582	2\\
50	5	0.116615355910066	2\\
50	10	0.157084511467169	2\\
75	1	0.0532821824381927	2\\
75	5	0.131648235913639	2\\
75	10	0.180271127776175	2\\
100	1	0.0559190292456523	2\\
100	5	0.138753989177189	2\\
100	10	0.200134089840193	2\\
150	1	0.0701754385964912	2\\
150	5	0.187798364651841	2\\
150	10	0.310488615934897	2\\
};
\addlegendentry{MmWave UM}

\addplot3[%
surf,
fill opacity=0.4, shader=flat corner, draw=black, z buffer=sort, colormap={mymap}{[1pt] rgb(0pt)=(0.2081,0.1663,0.5292); rgb(1pt)=(0.211624,0.189781,0.577676); rgb(2pt)=(0.212252,0.213771,0.626971); rgb(3pt)=(0.2081,0.2386,0.677086); rgb(4pt)=(0.195905,0.264457,0.7279); rgb(5pt)=(0.170729,0.291938,0.779248); rgb(6pt)=(0.125271,0.324243,0.830271); rgb(7pt)=(0.0591333,0.359833,0.868333); rgb(8pt)=(0.0116952,0.38751,0.881957); rgb(9pt)=(0.00595714,0.408614,0.882843); rgb(10pt)=(0.0165143,0.4266,0.878633); rgb(11pt)=(0.0328524,0.443043,0.871957); rgb(12pt)=(0.0498143,0.458571,0.864057); rgb(13pt)=(0.0629333,0.47369,0.855438); rgb(14pt)=(0.0722667,0.488667,0.8467); rgb(15pt)=(0.0779429,0.503986,0.838371); rgb(16pt)=(0.0793476,0.520024,0.831181); rgb(17pt)=(0.0749429,0.537543,0.826271); rgb(18pt)=(0.0640571,0.556986,0.823957); rgb(19pt)=(0.0487714,0.577224,0.822829); rgb(20pt)=(0.0343429,0.596581,0.819852); rgb(21pt)=(0.0265,0.6137,0.8135); rgb(22pt)=(0.0238905,0.628662,0.803762); rgb(23pt)=(0.0230905,0.641786,0.791267); rgb(24pt)=(0.0227714,0.653486,0.776757); rgb(25pt)=(0.0266619,0.664195,0.760719); rgb(26pt)=(0.0383714,0.674271,0.743552); rgb(27pt)=(0.0589714,0.683757,0.725386); rgb(28pt)=(0.0843,0.692833,0.706167); rgb(29pt)=(0.113295,0.7015,0.685857); rgb(30pt)=(0.145271,0.709757,0.664629); rgb(31pt)=(0.180133,0.717657,0.642433); rgb(32pt)=(0.217829,0.725043,0.619262); rgb(33pt)=(0.258643,0.731714,0.595429); rgb(34pt)=(0.302171,0.737605,0.571186); rgb(35pt)=(0.348167,0.742433,0.547267); rgb(36pt)=(0.395257,0.7459,0.524443); rgb(37pt)=(0.44201,0.748081,0.503314); rgb(38pt)=(0.487124,0.749062,0.483976); rgb(39pt)=(0.530029,0.749114,0.466114); rgb(40pt)=(0.570857,0.748519,0.44939); rgb(41pt)=(0.609852,0.747314,0.433686); rgb(42pt)=(0.6473,0.7456,0.4188); rgb(43pt)=(0.683419,0.743476,0.404433); rgb(44pt)=(0.71841,0.741133,0.390476); rgb(45pt)=(0.752486,0.7384,0.376814); rgb(46pt)=(0.785843,0.735567,0.363271); rgb(47pt)=(0.818505,0.732733,0.34979); rgb(48pt)=(0.850657,0.7299,0.336029); rgb(49pt)=(0.882433,0.727433,0.3217); rgb(50pt)=(0.913933,0.725786,0.306276); rgb(51pt)=(0.944957,0.726114,0.288643); rgb(52pt)=(0.973895,0.731395,0.266648); rgb(53pt)=(0.993771,0.745457,0.240348); rgb(54pt)=(0.999043,0.765314,0.216414); rgb(55pt)=(0.995533,0.786057,0.196652); rgb(56pt)=(0.988,0.8066,0.179367); rgb(57pt)=(0.978857,0.827143,0.163314); rgb(58pt)=(0.9697,0.848138,0.147452); rgb(59pt)=(0.962586,0.870514,0.1309); rgb(60pt)=(0.958871,0.8949,0.113243); rgb(61pt)=(0.959824,0.921833,0.0948381); rgb(62pt)=(0.9661,0.951443,0.0755333); rgb(63pt)=(0.9763,0.9831,0.0538)}, mesh/rows=4,area legend,fill=blue]
table[row sep=crcr, point meta=\thisrow{c}] {%
x	y	z	c\\
50	1	0.0316896945113449	1\\
50	5	0.0675201209960568	1\\
50	10	0.0961316619241713	1\\
75	1	0.0431480842250604	1\\
75	5	0.0970195591431233	1\\
75	10	0.136754006892402	1\\
100	1	0.0407955957074874	1\\
100	5	0.0882316610492509	1\\
100	10	0.130161725944486	1\\
150	1	0.0623124783853591	1\\
150	5	0.153172430800525	1\\
150	10	0.250475904218014	1\\
};
\addlegendentry{MmWave AM}

\end{axis}
\end{tikzpicture}%

%% file: figures/dist-mptcp.tex
%
%
\begin{tikzpicture}
\pgfplotsset{every tick label/.append style={font=\scriptsize}}

\begin{axis}[%
width=0.951\fwidth,
height=\fheight,
at={(0\fwidth,0\fheight)},
scale only axis,
xmin=45,
xmax=155,
xtick={ 50,  75, 100, 150},
xlabel style={font=\scriptsize\color{white!15!black}},
xlabel={Distance from eNB [m]},
ymin=0,
ymax=4000,
ylabel style={font=\scriptsize\color{white!15!black}},
ylabel={Throughput [Mbit/s]},
axis background/.style={fill=white},
xmajorgrids,
ymajorgrids,
ylabel shift = -5 pt,
yticklabel shift = -2 pt,
legend style={font=\scriptsize, legend cell align=left, align=left, draw=white!15!black}
]
\addplot [color=blue, dashed, mark=x, mark options={solid, blue}]
  table[row sep=crcr]{%
50	2838.95204081818\\
75	2027.12036676364\\
100	1806.275942\\
150	875.598794381394\\
};
\addlegendentry{LTE + 28 GHz mmWave, CUBIC}

\addplot [color=red, dashed, mark=o, mark options={solid, red}]
  table[row sep=crcr]{%
50	1466.24443354545\\
75	1064.86984872727\\
100	986.925018377359\\
150	540.342533475675\\
};
\addlegendentry{LTE + 28 GHz mmWave, BALIA}

\addplot [color=blue, mark=triangle, mark options={solid, rotate=180, blue}]
  table[row sep=crcr]{%
50	3881.91622190909\\
75	1913.90920686275\\
100	1318.25001145283\\
150	629.220500463768\\
};
\addlegendentry{28 GHz + 73 GHz mmWave, CUBIC}

\addplot [color=red, mark=triangle, mark options={solid, red}]
  table[row sep=crcr]{%
50	3653.00844454546\\
75	1428.82549348936\\
100	679.689223018868\\
150	511.463499198895\\
};
\addlegendentry{28 GHz + 73 GHz mmWave, BALIA}

\addplot [color=black, dotted, line width=0.7pt, mark=square, mark options={solid, black}]
  table[row sep=crcr]{%
50	2843.81612863636\\
75	1582.09386244156\\
100	782.847855418182\\
150	424.976978755\\
};
\addlegendentry{28 GHz mmWave, CUBIC}

\end{axis}
\end{tikzpicture}%

%% file: figures/dist-cubic-ratio-lte.tex
%
%
\definecolor{mycolor1}{rgb}{0.05913,0.35983,0.86833}%
\definecolor{mycolor2}{rgb}{0.07227,0.48867,0.84670}%
\definecolor{mycolor3}{rgb}{0.02650,0.61370,0.81350}%
\definecolor{mycolor4}{rgb}{0.08430,0.69283,0.70617}%
\definecolor{mycolor5}{rgb}{0.34817,0.74243,0.54727}%
\definecolor{mycolor6}{rgb}{0.64730,0.74560,0.41880}%
\definecolor{mycolor7}{rgb}{0.88243,0.72743,0.32170}%
\definecolor{mycolor8}{rgb}{0.98800,0.80660,0.17937}%
\begin{tikzpicture}
\pgfplotsset{every tick label/.append style={font=\scriptsize}}
\begin{axis}[%
ybar,
width=\fwidth,
height=\fheight,
at={(0\fwidth,0\fheight)},
scale only axis,
bar shift auto,
xtick=data,
enlarge x limits=0.4,
bar width=15pt,
symbolic x coords={100,150},
xlabel style={font=\scriptsize\color{white!15!black}},
xlabel={Distance from eNB [m]},
ymin=0,
ymax=1800,
ylabel style={font=\scriptsize\color{white!15!black}},
ylabel={Throughput [Mbit/s]},
axis background/.style={fill=white},
xmajorgrids,
ymajorgrids,
ylabel shift = -5 pt,
yticklabel shift = -2 pt,
legend style={font=\scriptsize, at={(0.16,0.5)}, anchor=south west, legend cell align=left, align=left, draw=white!15!black}
]
\addplot[fill=mycolor1,postaction={pattern=crosshatch dots}] coordinates {(100,986.741805183363) (150,427.076948032215)};
\addplot[fill=mycolor3] coordinates {(100,331.508206269468) (150,202.143552431553)};
\addplot[fill=mycolor5,postaction={pattern=crosshatch dots}] coordinates {(100,1734.39645657742) (150,803.615063405256)};
\addplot[fill=mycolor7] coordinates {(100,71.8794854225764) (150,71.9837309761383)};
\legend{mmWave 28 GHz (second subflow on mmWave), mmWave 73 GHz, mmWave 28 GHz (second subflow on LTE), LTE}
\end{axis}
\end{tikzpicture}%

%% file: figures/wget-mptcp-cubic.tex
%
%
\begin{tikzpicture}
\pgfplotsset{every tick label/.append style={font=\scriptsize}}

\begin{axis}[%
width=\fwidth,
height=\fheight,
at={(0.772in,0.516in)},
scale only axis,
xmin=50,
xmax=150,
tick align=outside,
xlabel style={font=\scriptsize\color{white!15!black}},
xlabel={Distance [m]},
ymin=0.1,
ymax=10,
ytick={0.1, 1, 5, 10},
ylabel style={font=\scriptsize\color{white!15!black}},
ylabel={Filesize [MB]},
zmin=0,
zmax=0.25,
zlabel style={font=\scriptsize\color{white!15!black}},
zlabel={Download time [s]},
view={-37.5}{30},
axis background/.style={fill=white},
axis x line*=bottom,
axis y line*=left,
axis z line*=left,
xmajorgrids,
ymajorgrids,
zmajorgrids,
legend style={font=\scriptsize, at={(0.45,0.68)}, anchor=south west, legend cell align=left, align=left, draw=white!15!black}
]

\addplot3[%
surf,
fill opacity=0.4, shader=flat corner, draw=black, z buffer=sort, colormap={mymap}{[1pt] rgb(0pt)=(0.2081,0.1663,0.5292); rgb(1pt)=(0.211624,0.189781,0.577676); rgb(2pt)=(0.212252,0.213771,0.626971); rgb(3pt)=(0.2081,0.2386,0.677086); rgb(4pt)=(0.195905,0.264457,0.7279); rgb(5pt)=(0.170729,0.291938,0.779248); rgb(6pt)=(0.125271,0.324243,0.830271); rgb(7pt)=(0.0591333,0.359833,0.868333); rgb(8pt)=(0.0116952,0.38751,0.881957); rgb(9pt)=(0.00595714,0.408614,0.882843); rgb(10pt)=(0.0165143,0.4266,0.878633); rgb(11pt)=(0.0328524,0.443043,0.871957); rgb(12pt)=(0.0498143,0.458571,0.864057); rgb(13pt)=(0.0629333,0.47369,0.855438); rgb(14pt)=(0.0722667,0.488667,0.8467); rgb(15pt)=(0.0779429,0.503986,0.838371); rgb(16pt)=(0.0793476,0.520024,0.831181); rgb(17pt)=(0.0749429,0.537543,0.826271); rgb(18pt)=(0.0640571,0.556986,0.823957); rgb(19pt)=(0.0487714,0.577224,0.822829); rgb(20pt)=(0.0343429,0.596581,0.819852); rgb(21pt)=(0.0265,0.6137,0.8135); rgb(22pt)=(0.0238905,0.628662,0.803762); rgb(23pt)=(0.0230905,0.641786,0.791267); rgb(24pt)=(0.0227714,0.653486,0.776757); rgb(25pt)=(0.0266619,0.664195,0.760719); rgb(26pt)=(0.0383714,0.674271,0.743552); rgb(27pt)=(0.0589714,0.683757,0.725386); rgb(28pt)=(0.0843,0.692833,0.706167); rgb(29pt)=(0.113295,0.7015,0.685857); rgb(30pt)=(0.145271,0.709757,0.664629); rgb(31pt)=(0.180133,0.717657,0.642433); rgb(32pt)=(0.217829,0.725043,0.619262); rgb(33pt)=(0.258643,0.731714,0.595429); rgb(34pt)=(0.302171,0.737605,0.571186); rgb(35pt)=(0.348167,0.742433,0.547267); rgb(36pt)=(0.395257,0.7459,0.524443); rgb(37pt)=(0.44201,0.748081,0.503314); rgb(38pt)=(0.487124,0.749062,0.483976); rgb(39pt)=(0.530029,0.749114,0.466114); rgb(40pt)=(0.570857,0.748519,0.44939); rgb(41pt)=(0.609852,0.747314,0.433686); rgb(42pt)=(0.6473,0.7456,0.4188); rgb(43pt)=(0.683419,0.743476,0.404433); rgb(44pt)=(0.71841,0.741133,0.390476); rgb(45pt)=(0.752486,0.7384,0.376814); rgb(46pt)=(0.785843,0.735567,0.363271); rgb(47pt)=(0.818505,0.732733,0.34979); rgb(48pt)=(0.850657,0.7299,0.336029); rgb(49pt)=(0.882433,0.727433,0.3217); rgb(50pt)=(0.913933,0.725786,0.306276); rgb(51pt)=(0.944957,0.726114,0.288643); rgb(52pt)=(0.973895,0.731395,0.266648); rgb(53pt)=(0.993771,0.745457,0.240348); rgb(54pt)=(0.999043,0.765314,0.216414); rgb(55pt)=(0.995533,0.786057,0.196652); rgb(56pt)=(0.988,0.8066,0.179367); rgb(57pt)=(0.978857,0.827143,0.163314); rgb(58pt)=(0.9697,0.848138,0.147452); rgb(59pt)=(0.962586,0.870514,0.1309); rgb(60pt)=(0.958871,0.8949,0.113243); rgb(61pt)=(0.959824,0.921833,0.0948381); rgb(62pt)=(0.9661,0.951443,0.0755333); rgb(63pt)=(0.9763,0.9831,0.0538)}, mesh/rows=4,area legend,fill=yellow]
table[row sep=crcr, point meta=\thisrow{c}] {%
x	y	z	c\\
50	0.1	0.00974354977005223	1\\
50	0.5	0.0227977384643443	1\\
50	1	0.0289687137891078	1\\
50	5	0.0586579070858752	1\\
50	10	0.0872052462676155	1\\
75	0.1	0.00984290720107091	1\\
75	0.5	0.027038719446247	1\\
75	1	0.0360334390314212	1\\
75	5	0.0790688848124486	1\\
75	10	0.11490554763984	1\\
100	0.1	0.0104038244458663	1\\
100	0.5	0.0315437511828907	1\\
100	1	0.0436357289348519	1\\
100	5	0.104412471025539	1\\
100	10	0.149309443822322	1\\
150	0.1	0.0126048328189511	1\\
150	0.5	0.042552105052637	1\\
150	1	0.0599279665841658	1\\
150	5	0.152975136950991	1\\
150	10	0.23128125187916	1\\
};
\addlegendentry{LTE + 28 GHz mmWave, CUBIC}

\addplot3[%
surf,
fill opacity=0.2, shader=flat corner, draw=black, z buffer=sort, colormap={mymap}{[1pt] rgb(0pt)=(0.2081,0.1663,0.5292); rgb(1pt)=(0.211624,0.189781,0.577676); rgb(2pt)=(0.212252,0.213771,0.626971); rgb(3pt)=(0.2081,0.2386,0.677086); rgb(4pt)=(0.195905,0.264457,0.7279); rgb(5pt)=(0.170729,0.291938,0.779248); rgb(6pt)=(0.125271,0.324243,0.830271); rgb(7pt)=(0.0591333,0.359833,0.868333); rgb(8pt)=(0.0116952,0.38751,0.881957); rgb(9pt)=(0.00595714,0.408614,0.882843); rgb(10pt)=(0.0165143,0.4266,0.878633); rgb(11pt)=(0.0328524,0.443043,0.871957); rgb(12pt)=(0.0498143,0.458571,0.864057); rgb(13pt)=(0.0629333,0.47369,0.855438); rgb(14pt)=(0.0722667,0.488667,0.8467); rgb(15pt)=(0.0779429,0.503986,0.838371); rgb(16pt)=(0.0793476,0.520024,0.831181); rgb(17pt)=(0.0749429,0.537543,0.826271); rgb(18pt)=(0.0640571,0.556986,0.823957); rgb(19pt)=(0.0487714,0.577224,0.822829); rgb(20pt)=(0.0343429,0.596581,0.819852); rgb(21pt)=(0.0265,0.6137,0.8135); rgb(22pt)=(0.0238905,0.628662,0.803762); rgb(23pt)=(0.0230905,0.641786,0.791267); rgb(24pt)=(0.0227714,0.653486,0.776757); rgb(25pt)=(0.0266619,0.664195,0.760719); rgb(26pt)=(0.0383714,0.674271,0.743552); rgb(27pt)=(0.0589714,0.683757,0.725386); rgb(28pt)=(0.0843,0.692833,0.706167); rgb(29pt)=(0.113295,0.7015,0.685857); rgb(30pt)=(0.145271,0.709757,0.664629); rgb(31pt)=(0.180133,0.717657,0.642433); rgb(32pt)=(0.217829,0.725043,0.619262); rgb(33pt)=(0.258643,0.731714,0.595429); rgb(34pt)=(0.302171,0.737605,0.571186); rgb(35pt)=(0.348167,0.742433,0.547267); rgb(36pt)=(0.395257,0.7459,0.524443); rgb(37pt)=(0.44201,0.748081,0.503314); rgb(38pt)=(0.487124,0.749062,0.483976); rgb(39pt)=(0.530029,0.749114,0.466114); rgb(40pt)=(0.570857,0.748519,0.44939); rgb(41pt)=(0.609852,0.747314,0.433686); rgb(42pt)=(0.6473,0.7456,0.4188); rgb(43pt)=(0.683419,0.743476,0.404433); rgb(44pt)=(0.71841,0.741133,0.390476); rgb(45pt)=(0.752486,0.7384,0.376814); rgb(46pt)=(0.785843,0.735567,0.363271); rgb(47pt)=(0.818505,0.732733,0.34979); rgb(48pt)=(0.850657,0.7299,0.336029); rgb(49pt)=(0.882433,0.727433,0.3217); rgb(50pt)=(0.913933,0.725786,0.306276); rgb(51pt)=(0.944957,0.726114,0.288643); rgb(52pt)=(0.973895,0.731395,0.266648); rgb(53pt)=(0.993771,0.745457,0.240348); rgb(54pt)=(0.999043,0.765314,0.216414); rgb(55pt)=(0.995533,0.786057,0.196652); rgb(56pt)=(0.988,0.8066,0.179367); rgb(57pt)=(0.978857,0.827143,0.163314); rgb(58pt)=(0.9697,0.848138,0.147452); rgb(59pt)=(0.962586,0.870514,0.1309); rgb(60pt)=(0.958871,0.8949,0.113243); rgb(61pt)=(0.959824,0.921833,0.0948381); rgb(62pt)=(0.9661,0.951443,0.0755333); rgb(63pt)=(0.9763,0.9831,0.0538)}, mesh/rows=4,area legend,fill=blue]
table[row sep=crcr, point meta=\thisrow{c}] {%
x	y	z	c\\
50	0.1	0.00699300699300699	3\\
50	0.5	0.017382839660687	3\\
50	1	0.0227025063567018	3\\
50	5	0.0412650204674502	3\\
50	10	0.0582031522827276	3\\
75	0.1	0.00799923207372092	3\\
75	0.5	0.020256036298817	3\\
75	1	0.0250551212667869	3\\
75	5	0.0535515380001714	3\\
75	10	0.0740455528240974	3\\
100	0.1	0.00903240374844756	3\\
100	0.5	0.0226437481743478	3\\
100	1	0.0294376091951316	3\\
100	5	0.0687152515562287	3\\
100	10	0.0973288107879254	3\\
150	0.1	0.0111803468604294	3\\
150	0.5	0.0355689967040588	3\\
150	1	0.0502228481150028	3\\
150	5	0.119627723334139	3\\
150	10	0.181188780499353	3\\
};
\addlegendentry{28 + 73 GHz mmWave, CUBIC}

\end{axis}
\end{tikzpicture}%

%% file: figures/olia-time.tex
\begin{tikzpicture}
\pgfplotsset{every tick label/.append style={font=\scriptsize}}
\begin{axis}[%
width=0.951\fwidth,
height=0.265\fheight,
at={(0\fwidth,0.735\fheight)},
scale only axis,
xmin=2,
xmax=12,
xlabel style={font=\scriptsize\color{white!15!black}},
xlabel={Time [s]},
ymin=0,
ymax=500,
ylabel style={font=\scriptsize\color{white!15!black}},
ylabel={Throughput [Mbit/s]},
axis background/.style={fill=white},
xmajorgrids,
ymajorgrids,
ylabel shift = -5 pt,
yticklabel shift = -2 pt,
legend style={font=\scriptsize,at={(0.608,0.661)}, anchor=south west, legend cell align=left, align=left, draw=white!15!black}
]
\addplot [color=black, dashed]
  table[row sep=crcr]{%
2	47.92704\\
2.25	75.268224\\
2.5	75.268224\\
2.75	75.316288\\
3	75.268224\\
3.25	75.268224\\
3.5	75.316288\\
3.75	75.270592\\
4	75.268224\\
4.25	75.316288\\
4.5	75.268224\\
4.75	75.268224\\
5	75.316288\\
5.25	75.268224\\
5.5	75.268224\\
5.75	75.316288\\
6	75.268224\\
6.25	74.066624\\
6.5	49.409792\\
6.75	75.268224\\
7	75.268224\\
7.25	75.318656\\
7.5	75.268224\\
7.75	75.268224\\
8	75.316288\\
8.25	75.268224\\
8.5	75.268224\\
8.75	75.316288\\
9	75.268224\\
9.25	75.268224\\
9.5	75.316288\\
9.75	75.268224\\
10	75.270592\\
10.25	75.268224\\
10.5	75.316288\\
10.75	75.268224\\
11	75.268224\\
11.25	75.316288\\
11.5	75.268224\\
11.75	75.268224\\
12	75.316288\\
12.25	75.268224\\
};
\addlegendentry{LTE subflow}

\addplot [color=red]
  table[row sep=crcr]{%
2	196.738688\\
2.25	244.117056\\
2.5	324.047488\\
2.75	314.915328\\
3	310.397312\\
3.25	303.04352\\
3.5	322.89632\\
3.75	223.689856\\
4	297.083584\\
4.25	169.18528\\
4.5	308.33056\\
4.75	232.725888\\
5	183.268032\\
5.25	255.508224\\
5.5	151.16128\\
5.75	136.792512\\
6	168.896896\\
6.25	337.072832\\
6.5	127.465728\\
6.75	0\\
7	0\\
7.25	48.977216\\
7.5	0\\
7.75	0\\
8	0\\
8.25	0\\
8.5	0\\
8.75	71.759552\\
9	205.764352\\
9.25	218.114432\\
9.5	0.769024\\
9.75	0\\
10	0\\
10.25	0\\
10.5	0\\
10.75	0\\
11	0\\
11.25	0\\
11.5	0\\
11.75	0\\
12	0\\
};
\addlegendentry{mmWave subflow}

\end{axis}
\end{tikzpicture}

%% file: figures/balia-time.tex
\begin{tikzpicture}
\pgfplotsset{every tick label/.append style={font=\scriptsize}}
\begin{axis}[%
width=0.951\fwidth,
height=0.265\fheight,
at={(0\fwidth,0.368\fheight)},
scale only axis,
xmin=2,
xmax=12,
xlabel style={font=\scriptsize\color{white!15!black}},
xlabel={Time [s]},
ymin=0,
ymax=500,
ylabel style={font=\scriptsize\color{white!15!black}},
ylabel={Throughput [Mbit/s]},
axis background/.style={fill=white},
xmajorgrids,
ymajorgrids,
ylabel shift = -5 pt,
yticklabel shift = -2 pt,
legend style={font=\scriptsize,at={(0.608,0.861)}, anchor=south west, legend cell align=left, align=left, draw=white!15!black}
]
\addplot [color=black, dashed]
  table[row sep=crcr]{%
2	47.92704\\
2.25	75.268224\\
2.5	75.268224\\
2.75	75.316288\\
3	75.268224\\
3.25	75.268224\\
3.5	75.316288\\
3.75	75.270592\\
4	75.268224\\
4.25	75.316288\\
4.5	75.268224\\
4.75	75.268224\\
5	75.316288\\
5.25	75.268224\\
5.5	75.268224\\
5.75	75.316288\\
6	75.268224\\
6.25	70.17344\\
6.5	45.756928\\
6.75	75.268224\\
7	75.316288\\
7.25	75.270592\\
7.5	75.268224\\
7.75	75.316288\\
8	75.268224\\
8.25	75.268224\\
8.5	75.316288\\
8.75	75.268224\\
9	75.268224\\
9.25	75.316288\\
9.5	75.270592\\
9.75	75.268224\\
10	75.316288\\
10.25	53.254912\\
10.5	0\\
10.75	0\\
11	0\\
11.25	34.894464\\
11.5	66.80896\\
11.75	75.268224\\
12	75.316288\\
12.25	75.268224\\
};
\addlegendentry{LTE subflow}

\addplot [color=red]
  table[row sep=crcr]{%
2	196.738688\\
2.25	244.117056\\
2.5	324.047488\\
2.75	314.915328\\
3	310.397312\\
3.25	303.04352\\
3.5	322.89632\\
3.75	223.689856\\
4	297.083584\\
4.25	169.18528\\
4.5	308.33056\\
4.75	232.725888\\
5	183.268032\\
5.25	255.652416\\
5.5	193.986304\\
5.75	123.622976\\
6	168.848832\\
6.25	308.33056\\
6.5	137.126592\\
6.75	0\\
7	0\\
7.25	25.666176\\
7.5	0\\
7.75	0\\
8	0\\
8.25	0\\
8.5	0\\
8.75	71.37504\\
9	222.058048\\
9.25	217.153152\\
9.5	253.153088\\
9.75	206.002304\\
10	353.943296\\
10.25	65.991872\\
10.5	53.447168\\
10.75	118.573888\\
11	134.483072\\
11.25	96.416384\\
11.5	280.020864\\
11.75	346.685632\\
12	115.257472\\
12.25	350.434624\\
12.5	284.250496\\
};
\addlegendentry{mmWave subflow}

\end{axis}
\end{tikzpicture}

%% file: figures/cubic-time.tex
\begin{tikzpicture}
\pgfplotsset{every tick label/.append style={font=\scriptsize}}
\begin{axis}[%
width=0.951\fwidth,
height=0.265\fheight,
at={(0\fwidth,0\fheight)},
scale only axis,
xmin=2,
xmax=12,
xlabel style={font=\scriptsize\color{white!15!black}},
xlabel={Time [s]},
ymin=0,
ymax=500,
ylabel style={font=\scriptsize\color{white!15!black}},
ylabel={Throughput [Mbit/s]},
axis background/.style={fill=white},
xmajorgrids,
ymajorgrids,
ylabel shift = -5 pt,
yticklabel shift = -2 pt,
legend style={font=\scriptsize,at={(0.608,0.861)}, anchor=south west, legend cell align=left, align=left, draw=white!15!black}
]
\addplot [color=black, dashed]
  table[row sep=crcr]{%
2	47.92704\\
2.25	75.268224\\
2.5	75.268224\\
2.75	75.316288\\
3	75.270592\\
3.25	75.268224\\
3.5	75.316288\\
3.75	75.268224\\
4	75.268224\\
4.25	75.316288\\
4.5	75.268224\\
4.75	75.268224\\
5	75.318656\\
5.25	75.268224\\
5.5	75.268224\\
5.75	75.316288\\
6	75.268224\\
6.25	75.268224\\
6.5	75.316288\\
6.75	75.268224\\
7	75.268224\\
7.25	75.316288\\
7.5	75.268224\\
7.75	75.268224\\
8	75.268224\\
8.25	75.316288\\
8.5	75.268224\\
8.75	75.268224\\
9	75.316288\\
9.25	75.270592\\
9.5	75.268224\\
9.75	75.316288\\
10	75.268224\\
10.25	75.268224\\
10.5	75.316288\\
10.75	75.268224\\
11	75.268224\\
11.25	75.316288\\
11.5	75.268224\\
11.75	75.268224\\
12	75.316288\\
12.25	75.268224\\
};
\addlegendentry{LTE subflow}

\addplot [color=red]
  table[row sep=crcr]{%
2	184.914944\\
2.25	232.966208\\
2.5	323.95136\\
2.75	314.915328\\
3	310.447744\\
3.25	311.262464\\
3.5	323.182336\\
3.75	225.324032\\
4	298.621632\\
4.25	173.559104\\
4.5	315.732416\\
4.75	237.340032\\
5	190.047424\\
5.25	258.488192\\
5.5	195.62048\\
5.75	127.898304\\
6	176.779392\\
6.25	337.16896\\
6.5	291.844608\\
6.75	363.796416\\
7	189.804736\\
7.25	303.331904\\
7.5	348.031424\\
7.75	171.011712\\
8	119.19872\\
8.25	157.4096\\
8.5	233.014272\\
8.75	362.979328\\
9	377.016384\\
9.25	226.237248\\
9.5	259.5456\\
9.75	211.72192\\
10	409.841728\\
10.25	274.397376\\
10.5	272.7632\\
10.75	219.123776\\
11	249.596352\\
11.25	148.75808\\
11.5	283.673728\\
11.75	346.204992\\
12	115.738112\\
12.25	355.721664\\
12.5	286.557568\\
};
\addlegendentry{mmWave subflow}
\end{axis}
\end{tikzpicture}

%% file: figures/wget-mptcp-lte-balia-cubic-small.tex
%
%
\begin{tikzpicture}
\pgfplotsset{every tick label/.append style={font=\scriptsize}}

\begin{axis}[%
width=\fwidth,
height=\fheight,
at={(0\fwidth,0\fheight)},
scale only axis,
xmin=50,
xmax=150,
tick align=outside,
xlabel style={font=\scriptsize\color{white!15!black}},
xlabel={Distance [m]},
ymin=0.1,
ymax=1,
ytick={0.1, 0.5, 1},
yticklabels={0.1, 0.5, 1},
ylabel style={font=\scriptsize\color{white!15!black}},
ylabel={Filesize [MB]},
zticklabel style={
        /pgf/number format/fixed,
        /pgf/number format/precision=5
},
scaled z ticks=false,
zmin=0,
zmax=0.06,
zlabel style={font=\scriptsize\color{white!15!black}},
zlabel={Download time [s]},
view={-37.5}{30},
axis background/.style={fill=white},
axis x line*=bottom,
axis y line*=left,
axis z line*=left,
xmajorgrids,
ymajorgrids,
zmajorgrids,
legend style={font=\scriptsize, at={(0.334,0.75)}, anchor=south west, legend cell align=left, align=left, draw=white!15!black}
]

\addplot3[%
surf,
fill opacity=0.4, shader=flat corner, draw=black, z buffer=sort, colormap={mymap}{[1pt] rgb(0pt)=(0.2081,0.1663,0.5292); rgb(1pt)=(0.211624,0.189781,0.577676); rgb(2pt)=(0.212252,0.213771,0.626971); rgb(3pt)=(0.2081,0.2386,0.677086); rgb(4pt)=(0.195905,0.264457,0.7279); rgb(5pt)=(0.170729,0.291938,0.779248); rgb(6pt)=(0.125271,0.324243,0.830271); rgb(7pt)=(0.0591333,0.359833,0.868333); rgb(8pt)=(0.0116952,0.38751,0.881957); rgb(9pt)=(0.00595714,0.408614,0.882843); rgb(10pt)=(0.0165143,0.4266,0.878633); rgb(11pt)=(0.0328524,0.443043,0.871957); rgb(12pt)=(0.0498143,0.458571,0.864057); rgb(13pt)=(0.0629333,0.47369,0.855438); rgb(14pt)=(0.0722667,0.488667,0.8467); rgb(15pt)=(0.0779429,0.503986,0.838371); rgb(16pt)=(0.0793476,0.520024,0.831181); rgb(17pt)=(0.0749429,0.537543,0.826271); rgb(18pt)=(0.0640571,0.556986,0.823957); rgb(19pt)=(0.0487714,0.577224,0.822829); rgb(20pt)=(0.0343429,0.596581,0.819852); rgb(21pt)=(0.0265,0.6137,0.8135); rgb(22pt)=(0.0238905,0.628662,0.803762); rgb(23pt)=(0.0230905,0.641786,0.791267); rgb(24pt)=(0.0227714,0.653486,0.776757); rgb(25pt)=(0.0266619,0.664195,0.760719); rgb(26pt)=(0.0383714,0.674271,0.743552); rgb(27pt)=(0.0589714,0.683757,0.725386); rgb(28pt)=(0.0843,0.692833,0.706167); rgb(29pt)=(0.113295,0.7015,0.685857); rgb(30pt)=(0.145271,0.709757,0.664629); rgb(31pt)=(0.180133,0.717657,0.642433); rgb(32pt)=(0.217829,0.725043,0.619262); rgb(33pt)=(0.258643,0.731714,0.595429); rgb(34pt)=(0.302171,0.737605,0.571186); rgb(35pt)=(0.348167,0.742433,0.547267); rgb(36pt)=(0.395257,0.7459,0.524443); rgb(37pt)=(0.44201,0.748081,0.503314); rgb(38pt)=(0.487124,0.749062,0.483976); rgb(39pt)=(0.530029,0.749114,0.466114); rgb(40pt)=(0.570857,0.748519,0.44939); rgb(41pt)=(0.609852,0.747314,0.433686); rgb(42pt)=(0.6473,0.7456,0.4188); rgb(43pt)=(0.683419,0.743476,0.404433); rgb(44pt)=(0.71841,0.741133,0.390476); rgb(45pt)=(0.752486,0.7384,0.376814); rgb(46pt)=(0.785843,0.735567,0.363271); rgb(47pt)=(0.818505,0.732733,0.34979); rgb(48pt)=(0.850657,0.7299,0.336029); rgb(49pt)=(0.882433,0.727433,0.3217); rgb(50pt)=(0.913933,0.725786,0.306276); rgb(51pt)=(0.944957,0.726114,0.288643); rgb(52pt)=(0.973895,0.731395,0.266648); rgb(53pt)=(0.993771,0.745457,0.240348); rgb(54pt)=(0.999043,0.765314,0.216414); rgb(55pt)=(0.995533,0.786057,0.196652); rgb(56pt)=(0.988,0.8066,0.179367); rgb(57pt)=(0.978857,0.827143,0.163314); rgb(58pt)=(0.9697,0.848138,0.147452); rgb(59pt)=(0.962586,0.870514,0.1309); rgb(60pt)=(0.958871,0.8949,0.113243); rgb(61pt)=(0.959824,0.921833,0.0948381); rgb(62pt)=(0.9661,0.951443,0.0755333); rgb(63pt)=(0.9763,0.9831,0.0538)}, mesh/rows=4, area legend,fill=yellow]
table[row sep=crcr, point meta=\thisrow{c}] {%
x	y	z	c\\
50	0.1	0.00974354977005223	1\\
50	0.5	0.0227977384643443	1\\
50	1	0.0289687137891078	1\\
75	0.1	0.00984290720107091	1\\
75	0.5	0.027038719446247	1\\
75	1	0.0360334390314212	1\\
100	0.1	0.0104038244458663	1\\
100	0.5	0.0315437511828907	1\\
100	1	0.0436357289348519	1\\
150	0.1	0.0126048328189511	1\\
150	0.5	0.042552105052637	1\\
150	1	0.0599279665841658	1\\
};
\addlegendentry{LTE + 28 GHz mmWave, CUBIC}

\addplot3[%
surf,
fill opacity=0.2, shader=flat corner, draw=black, z buffer=sort, colormap={mymap}{[1pt] rgb(0pt)=(0.2081,0.1663,0.5292); rgb(1pt)=(0.211624,0.189781,0.577676); rgb(2pt)=(0.212252,0.213771,0.626971); rgb(3pt)=(0.2081,0.2386,0.677086); rgb(4pt)=(0.195905,0.264457,0.7279); rgb(5pt)=(0.170729,0.291938,0.779248); rgb(6pt)=(0.125271,0.324243,0.830271); rgb(7pt)=(0.0591333,0.359833,0.868333); rgb(8pt)=(0.0116952,0.38751,0.881957); rgb(9pt)=(0.00595714,0.408614,0.882843); rgb(10pt)=(0.0165143,0.4266,0.878633); rgb(11pt)=(0.0328524,0.443043,0.871957); rgb(12pt)=(0.0498143,0.458571,0.864057); rgb(13pt)=(0.0629333,0.47369,0.855438); rgb(14pt)=(0.0722667,0.488667,0.8467); rgb(15pt)=(0.0779429,0.503986,0.838371); rgb(16pt)=(0.0793476,0.520024,0.831181); rgb(17pt)=(0.0749429,0.537543,0.826271); rgb(18pt)=(0.0640571,0.556986,0.823957); rgb(19pt)=(0.0487714,0.577224,0.822829); rgb(20pt)=(0.0343429,0.596581,0.819852); rgb(21pt)=(0.0265,0.6137,0.8135); rgb(22pt)=(0.0238905,0.628662,0.803762); rgb(23pt)=(0.0230905,0.641786,0.791267); rgb(24pt)=(0.0227714,0.653486,0.776757); rgb(25pt)=(0.0266619,0.664195,0.760719); rgb(26pt)=(0.0383714,0.674271,0.743552); rgb(27pt)=(0.0589714,0.683757,0.725386); rgb(28pt)=(0.0843,0.692833,0.706167); rgb(29pt)=(0.113295,0.7015,0.685857); rgb(30pt)=(0.145271,0.709757,0.664629); rgb(31pt)=(0.180133,0.717657,0.642433); rgb(32pt)=(0.217829,0.725043,0.619262); rgb(33pt)=(0.258643,0.731714,0.595429); rgb(34pt)=(0.302171,0.737605,0.571186); rgb(35pt)=(0.348167,0.742433,0.547267); rgb(36pt)=(0.395257,0.7459,0.524443); rgb(37pt)=(0.44201,0.748081,0.503314); rgb(38pt)=(0.487124,0.749062,0.483976); rgb(39pt)=(0.530029,0.749114,0.466114); rgb(40pt)=(0.570857,0.748519,0.44939); rgb(41pt)=(0.609852,0.747314,0.433686); rgb(42pt)=(0.6473,0.7456,0.4188); rgb(43pt)=(0.683419,0.743476,0.404433); rgb(44pt)=(0.71841,0.741133,0.390476); rgb(45pt)=(0.752486,0.7384,0.376814); rgb(46pt)=(0.785843,0.735567,0.363271); rgb(47pt)=(0.818505,0.732733,0.34979); rgb(48pt)=(0.850657,0.7299,0.336029); rgb(49pt)=(0.882433,0.727433,0.3217); rgb(50pt)=(0.913933,0.725786,0.306276); rgb(51pt)=(0.944957,0.726114,0.288643); rgb(52pt)=(0.973895,0.731395,0.266648); rgb(53pt)=(0.993771,0.745457,0.240348); rgb(54pt)=(0.999043,0.765314,0.216414); rgb(55pt)=(0.995533,0.786057,0.196652); rgb(56pt)=(0.988,0.8066,0.179367); rgb(57pt)=(0.978857,0.827143,0.163314); rgb(58pt)=(0.9697,0.848138,0.147452); rgb(59pt)=(0.962586,0.870514,0.1309); rgb(60pt)=(0.958871,0.8949,0.113243); rgb(61pt)=(0.959824,0.921833,0.0948381); rgb(62pt)=(0.9661,0.951443,0.0755333); rgb(63pt)=(0.9763,0.9831,0.0538)}, mesh/rows=4, area legend,fill=blue]
table[row sep=crcr, point meta=\thisrow{c}] {%
x	y	z	c\\
50	0.1	0.00970873786407767	3\\
50	0.5	0.0166711122966124	3\\
50	1	0.0249525900788502	3\\
75	0.1	0.00970873786407767	3\\
75	0.5	0.0175957207207207	3\\
75	1	0.029582297952905	3\\
100	0.1	0.00973662431235091	3\\
100	0.5	0.0189577053593433	3\\
100	1	0.0367080243741282	3\\
150	0.1	0.0102315240417794	3\\
150	0.5	0.0289212099477394	3\\
150	1	0.0514880032952322	3\\
};
\addlegendentry{LTE + 28 GHz mmWave, BALIA}

\end{axis}
\end{tikzpicture}%

%% file: figures/wget-mptcp-lte-balia-cubic-large.tex
%
%
\begin{tikzpicture}
\pgfplotsset{every tick label/.append style={font=\scriptsize}}

\begin{axis}[%
width=\fwidth,
height=\fheight,
at={(0\fwidth,0\fheight)},
scale only axis,
xmin=50,
xmax=150,
tick align=outside,
xlabel style={font=\scriptsize\color{white!15!black}},
xlabel={Distance [m]},
ymin=5,
ymax=10,
ylabel style={font=\scriptsize\color{white!15!black}},
ylabel={Filesize [MB]},
zmin=0,
zmax=0.78,
zlabel style={font=\scriptsize\color{white!15!black}},
zlabel={Download time [s]},
view={-37.5}{30},
axis background/.style={fill=white},
axis x line*=bottom,
axis y line*=left,
axis z line*=left,
xmajorgrids,
ymajorgrids,
zmajorgrids,
legend style={font=\scriptsize, at={(0.434,0.739)}, anchor=south west, legend cell align=left, align=left, draw=white!15!black}
]

\addplot3[%
surf,
fill opacity=0.4, shader=flat corner, draw=black, z buffer=sort, colormap={mymap}{[1pt] rgb(0pt)=(0.2081,0.1663,0.5292); rgb(1pt)=(0.211624,0.189781,0.577676); rgb(2pt)=(0.212252,0.213771,0.626971); rgb(3pt)=(0.2081,0.2386,0.677086); rgb(4pt)=(0.195905,0.264457,0.7279); rgb(5pt)=(0.170729,0.291938,0.779248); rgb(6pt)=(0.125271,0.324243,0.830271); rgb(7pt)=(0.0591333,0.359833,0.868333); rgb(8pt)=(0.0116952,0.38751,0.881957); rgb(9pt)=(0.00595714,0.408614,0.882843); rgb(10pt)=(0.0165143,0.4266,0.878633); rgb(11pt)=(0.0328524,0.443043,0.871957); rgb(12pt)=(0.0498143,0.458571,0.864057); rgb(13pt)=(0.0629333,0.47369,0.855438); rgb(14pt)=(0.0722667,0.488667,0.8467); rgb(15pt)=(0.0779429,0.503986,0.838371); rgb(16pt)=(0.0793476,0.520024,0.831181); rgb(17pt)=(0.0749429,0.537543,0.826271); rgb(18pt)=(0.0640571,0.556986,0.823957); rgb(19pt)=(0.0487714,0.577224,0.822829); rgb(20pt)=(0.0343429,0.596581,0.819852); rgb(21pt)=(0.0265,0.6137,0.8135); rgb(22pt)=(0.0238905,0.628662,0.803762); rgb(23pt)=(0.0230905,0.641786,0.791267); rgb(24pt)=(0.0227714,0.653486,0.776757); rgb(25pt)=(0.0266619,0.664195,0.760719); rgb(26pt)=(0.0383714,0.674271,0.743552); rgb(27pt)=(0.0589714,0.683757,0.725386); rgb(28pt)=(0.0843,0.692833,0.706167); rgb(29pt)=(0.113295,0.7015,0.685857); rgb(30pt)=(0.145271,0.709757,0.664629); rgb(31pt)=(0.180133,0.717657,0.642433); rgb(32pt)=(0.217829,0.725043,0.619262); rgb(33pt)=(0.258643,0.731714,0.595429); rgb(34pt)=(0.302171,0.737605,0.571186); rgb(35pt)=(0.348167,0.742433,0.547267); rgb(36pt)=(0.395257,0.7459,0.524443); rgb(37pt)=(0.44201,0.748081,0.503314); rgb(38pt)=(0.487124,0.749062,0.483976); rgb(39pt)=(0.530029,0.749114,0.466114); rgb(40pt)=(0.570857,0.748519,0.44939); rgb(41pt)=(0.609852,0.747314,0.433686); rgb(42pt)=(0.6473,0.7456,0.4188); rgb(43pt)=(0.683419,0.743476,0.404433); rgb(44pt)=(0.71841,0.741133,0.390476); rgb(45pt)=(0.752486,0.7384,0.376814); rgb(46pt)=(0.785843,0.735567,0.363271); rgb(47pt)=(0.818505,0.732733,0.34979); rgb(48pt)=(0.850657,0.7299,0.336029); rgb(49pt)=(0.882433,0.727433,0.3217); rgb(50pt)=(0.913933,0.725786,0.306276); rgb(51pt)=(0.944957,0.726114,0.288643); rgb(52pt)=(0.973895,0.731395,0.266648); rgb(53pt)=(0.993771,0.745457,0.240348); rgb(54pt)=(0.999043,0.765314,0.216414); rgb(55pt)=(0.995533,0.786057,0.196652); rgb(56pt)=(0.988,0.8066,0.179367); rgb(57pt)=(0.978857,0.827143,0.163314); rgb(58pt)=(0.9697,0.848138,0.147452); rgb(59pt)=(0.962586,0.870514,0.1309); rgb(60pt)=(0.958871,0.8949,0.113243); rgb(61pt)=(0.959824,0.921833,0.0948381); rgb(62pt)=(0.9661,0.951443,0.0755333); rgb(63pt)=(0.9763,0.9831,0.0538)}, mesh/rows=4,area legend,fill=yellow]
table[row sep=crcr, point meta=\thisrow{c}] {%
x	y	z	c\\
50	5	0.0586579070858752	1\\
50	10	0.0872052462676155	1\\
75	5	0.0790688848124486	1\\
75	10	0.11490554763984	1\\
100	5	0.104412471025539	1\\
100	10	0.149309443822322	1\\
150	5	0.152975136950991	1\\
150	10	0.23128125187916	1\\
};
\addlegendentry{LTE + 28 GHz mmWave, CUBIC}

\addplot3[%
surf,
fill opacity=0.2, shader=flat corner, draw=black, z buffer=sort, colormap={mymap}{[1pt] rgb(0pt)=(0.2081,0.1663,0.5292); rgb(1pt)=(0.211624,0.189781,0.577676); rgb(2pt)=(0.212252,0.213771,0.626971); rgb(3pt)=(0.2081,0.2386,0.677086); rgb(4pt)=(0.195905,0.264457,0.7279); rgb(5pt)=(0.170729,0.291938,0.779248); rgb(6pt)=(0.125271,0.324243,0.830271); rgb(7pt)=(0.0591333,0.359833,0.868333); rgb(8pt)=(0.0116952,0.38751,0.881957); rgb(9pt)=(0.00595714,0.408614,0.882843); rgb(10pt)=(0.0165143,0.4266,0.878633); rgb(11pt)=(0.0328524,0.443043,0.871957); rgb(12pt)=(0.0498143,0.458571,0.864057); rgb(13pt)=(0.0629333,0.47369,0.855438); rgb(14pt)=(0.0722667,0.488667,0.8467); rgb(15pt)=(0.0779429,0.503986,0.838371); rgb(16pt)=(0.0793476,0.520024,0.831181); rgb(17pt)=(0.0749429,0.537543,0.826271); rgb(18pt)=(0.0640571,0.556986,0.823957); rgb(19pt)=(0.0487714,0.577224,0.822829); rgb(20pt)=(0.0343429,0.596581,0.819852); rgb(21pt)=(0.0265,0.6137,0.8135); rgb(22pt)=(0.0238905,0.628662,0.803762); rgb(23pt)=(0.0230905,0.641786,0.791267); rgb(24pt)=(0.0227714,0.653486,0.776757); rgb(25pt)=(0.0266619,0.664195,0.760719); rgb(26pt)=(0.0383714,0.674271,0.743552); rgb(27pt)=(0.0589714,0.683757,0.725386); rgb(28pt)=(0.0843,0.692833,0.706167); rgb(29pt)=(0.113295,0.7015,0.685857); rgb(30pt)=(0.145271,0.709757,0.664629); rgb(31pt)=(0.180133,0.717657,0.642433); rgb(32pt)=(0.217829,0.725043,0.619262); rgb(33pt)=(0.258643,0.731714,0.595429); rgb(34pt)=(0.302171,0.737605,0.571186); rgb(35pt)=(0.348167,0.742433,0.547267); rgb(36pt)=(0.395257,0.7459,0.524443); rgb(37pt)=(0.44201,0.748081,0.503314); rgb(38pt)=(0.487124,0.749062,0.483976); rgb(39pt)=(0.530029,0.749114,0.466114); rgb(40pt)=(0.570857,0.748519,0.44939); rgb(41pt)=(0.609852,0.747314,0.433686); rgb(42pt)=(0.6473,0.7456,0.4188); rgb(43pt)=(0.683419,0.743476,0.404433); rgb(44pt)=(0.71841,0.741133,0.390476); rgb(45pt)=(0.752486,0.7384,0.376814); rgb(46pt)=(0.785843,0.735567,0.363271); rgb(47pt)=(0.818505,0.732733,0.34979); rgb(48pt)=(0.850657,0.7299,0.336029); rgb(49pt)=(0.882433,0.727433,0.3217); rgb(50pt)=(0.913933,0.725786,0.306276); rgb(51pt)=(0.944957,0.726114,0.288643); rgb(52pt)=(0.973895,0.731395,0.266648); rgb(53pt)=(0.993771,0.745457,0.240348); rgb(54pt)=(0.999043,0.765314,0.216414); rgb(55pt)=(0.995533,0.786057,0.196652); rgb(56pt)=(0.988,0.8066,0.179367); rgb(57pt)=(0.978857,0.827143,0.163314); rgb(58pt)=(0.9697,0.848138,0.147452); rgb(59pt)=(0.962586,0.870514,0.1309); rgb(60pt)=(0.958871,0.8949,0.113243); rgb(61pt)=(0.959824,0.921833,0.0948381); rgb(62pt)=(0.9661,0.951443,0.0755333); rgb(63pt)=(0.9763,0.9831,0.0538)}, mesh/rows=4,area legend,fill=blue]
table[row sep=crcr, point meta=\thisrow{c}] {%
x	y	z	c\\
50	5	0.275027502750275	3\\
50	10	0.603427468018344	3\\
75	5	0.303692905733722	3\\
75	10	0.65359477124183	3\\
100	5	0.333503420077573	3\\
100	10	0.693115972164463	3\\
150	5	0.382033718295977	3\\
150	10	0.744496311020779	3\\
};
\addlegendentry{LTE + 28 GHz mmWave, BALIA}

\end{axis}
\end{tikzpicture}%

%% file: figures/throughput-buffer-speed.tex
%
%
\definecolor{mycolor1}{rgb}{0.05913,0.35983,0.86833}%
\definecolor{mycolor2}{rgb}{0.98800,0.80660,0.17937}%
\definecolor{mycolor3}{rgb}{0.00000,0.70000,0.70000}%
\begin{tikzpicture}
\pgfplotsset{every tick label/.append style={font=\scriptsize}}

\begin{axis}[%
ybar,
width=\fwidth,
height=\fheight,
at={(0\fwidth,0\fheight)},
scale only axis,
bar shift auto,
xtick=data,
enlarge x limits=0.4,
bar width=15pt,
xlabel style={font=\scriptsize\color{white!15!black}},
xlabel={UE speed s [m/s]},
xticklabels={2,5},
ymin=0,
ymax=600,
ylabel style={font=\scriptsize\color{white!15!black}},
ylabel={Throughput [Mbit/s]},
axis background/.style={fill=white},
xmajorgrids,
ymajorgrids,
ylabel shift = -5 pt,
xlabel shift = -8 pt,
yticklabel shift = -2 pt,
xticklabel shift = -2 pt,
legend style={font=\scriptsize,at={(0.325,0.728)}, anchor=south west, legend cell align=left, align=left, draw=white!15!black}
]
\addplot[fill=mycolor1] coordinates {(1,401.242879618163) (2,382.910448045878)};
\addplot[fill=mycolor2] coordinates {(1,549.377191900724) (2,523.341909753792)};
\legend{RLC buffer size 2 MB, RLC buffer size 20 MB};
\draw[fill=mycolor3,postaction={pattern=crosshatch dots}] (0.89, 0) rectangle (0.96, 414.118292222952);
\draw[fill=mycolor3,postaction={pattern=crosshatch dots}] (1.04, 0) rectangle (1.11,
523.050648224);
\draw[fill=mycolor3,postaction={pattern=crosshatch dots}] (1.89, 0) rectangle (1.96, 405.318095504961);
\draw[fill=mycolor3,postaction={pattern=crosshatch dots}] (2.04, 0) rectangle (2.11, 508.372486191043);
\end{axis}
\end{tikzpicture}%